\documentclass[prx,twocolumn,floatfix,superscriptaddress,longbibliography,notitlepage, nofootinbib]{revtex4-1}

\usepackage{graphicx, color, graphpap}
\usepackage{enumitem}
\usepackage{amssymb}
\usepackage{amsthm}
\usepackage{dsfont}
\usepackage{mathtools}
\usepackage[dvipsnames]{xcolor}
\usepackage{multirow}
\usepackage[colorlinks=true,citecolor=blue,linkcolor=magenta]{hyperref}
\usepackage[T1]{fontenc}
\usepackage{bbm}
\usepackage{thmtools,thm-restate}
\usepackage{verbatim}
\usepackage{mathtools}
\usepackage{titlesec}
\usepackage{amsmath}

\usepackage[caption = false]{subfig}

\long\def\ca#1\cb{} 

\newcommand{\ket}[1]{|#1\rangle}               
\newcommand{\bra}[1]{\langle #1|}              

\def\I{ \mathbbm{1} }

\newcommand{\AC}{\mathcal{A}}

\newcommand{\OC}{\mathcal{O}}

\newcommand{\UC}{\mathcal{U}}

\newcommand{\Tr}{{\rm Tr}}

\newcommand{\Var}{{\rm Var}}

\renewcommand{\geq}{\geqslant}
\renewcommand{\leq}{\leqslant}

\DeclareMathOperator*{\argmin}{arg\,min}
\renewcommand{\vec}[1]{\boldsymbol{#1}}  

\newcommand*{\id}{\openone}

\newcommand{\thv}{\vec{\theta}}

\def\eRd{ { \varepsilon_{\scriptscriptstyle R}^{\scriptscriptstyle \diamond} } }
\def\eLd{ { \varepsilon_{\scriptscriptstyle L}^{\scriptscriptstyle \diamond} } }

   \theoremstyle{plain}
\newtheorem{thm}{Theorem}

\theoremstyle{definition}

\theoremstyle{remark}

{}
{}

\begin{document}

\title{Connecting ansatz expressibility to gradient magnitudes and barren plateaus}

\author{Zo\"{e} Holmes}
\affiliation{Information Sciences, Los Alamos National Laboratory, Los Alamos, NM, USA.}

\author{Kunal~Sharma} 
\affiliation{Hearne Institute for Theoretical Physics, Department of Physics and Astronomy, and Center for Computation and Technology,
Louisiana State University, Baton Rouge, Louisiana 70803, USA}
\affiliation{Theoretical Division, Los Alamos National Laboratory, Los Alamos, NM 87545, USA}

\author{M. Cerezo}
\affiliation{Theoretical Division, Los Alamos National Laboratory, Los Alamos, NM 87545, USA}
\affiliation{Center for Nonlinear Studies, Los Alamos National Laboratory, Los Alamos, NM, USA
}

\author{Patrick J. Coles}
\affiliation{Theoretical Division, Los Alamos National Laboratory, Los Alamos, NM 87545, USA}

\begin{abstract}
Parameterized quantum circuits serve as ans\"{a}tze for solving variational problems and provide a flexible paradigm for programming near-term quantum computers. Ideally, such ans\"{a}tze should be highly expressive so that a close approximation of the desired solution can be accessed. On the other hand, the ansatz must also have sufficiently large gradients to allow for training. Here, we derive a fundamental relationship between these two essential properties: expressibility and trainability. This is done by extending the well established barren plateau phenomenon, which holds for ans\"{a}tze that form exact 2-designs, to arbitrary ans\"{a}tze. Specifically, we calculate the variance in the cost gradient in terms of the expressibility of the ansatz, as measured by its distance from being a 2-design. Our resulting bounds indicate that highly expressive ans\"{a}tze exhibit flatter cost landscapes and therefore will be harder to train. Furthermore, we provide numerics illustrating the effect of expressiblity  on gradient scalings, and we discuss the implications for designing strategies to avoid barren plateaus.
\end{abstract}
 
\maketitle

\section{Introduction}

While quantum hardware is rapidly reaching the stage where it can outperform classical supercomputers~\cite{GoogleSupremacy2019}, we remain in the Noisy Intermediate-Scale Quantum (NISQ) era in which the available devices are relatively small and prone to errors~\cite{Preskill2018NISQ}. Variational quantum algorithms have gathered attention as a computational strategy that is well suited to the constraints imposed by NISQ devices~\cite{VQE,mcclean2016theory,qaoa2014,Romero,khatri2019quantum,VQSD,arrasmith2019variational,cerezo2020variationalfidelity,sharma2020noise,bravo-prieto2019,cerezo2020variational,heya2019subspace,cirstoiu2020variational,commeau2020variational,li2017efficient,endo2020variational,yuan2019theory,cerezo2020variationalreview}. In VQAs a problem-specific cost function is efficiently evaluated on a quantum computer, while a classical optimizer trains a parameterized quantum circuit to minimize this cost. The benefit of this paradigm is that it adapts to the qubit and connectivity constraints of NISQ devices, while keeping the circuit depth short to mitigate quantum hardware noise.

Central to the success of VQAs is the construction of a parameterized quantum circuit, which serves as an ansatz with which to explore the space of solutions to the target problem. Some noteworthy ans\"{a}tze include the quantum alternating operator ansatz~\cite{qaoa2014, hadfield2019quantum}, coupled cluster ansatz~\cite{bartlett2007coupled,lee2018generalized,cao2019quantum}, Hamiltonian variational ansatz~\cite{wecker2015progress}, and hardware efficient ansatz~\cite{kandala2017hardware}. To successfully find an optimal solution, the ansatz should ideally be both expressive and trainable. Specifically, the ansatz must be sufficiently expressive such that it contains a circuit that well-approximates the optimal solution. Concurrently, the cost landscape must be sufficiently featured to be able to train the parameters to find this optimal solution. 

Recently, it was shown that VQAs can exhibit barren plateaus, where under certain conditions the gradient of the cost function vanishes exponentially with the size of the system~\cite{mcclean2018barren,cerezo2020cost,sharma2020trainability,wang2020noise,cerezo2020impact,holmes2020barren,marrero2020entanglement,uvarov2020barren,arrasmith2020effect,abbas2020power}. In particular, Ref.~\cite{mcclean2018barren} demonstrated that if an ansatz is sufficiently random that it matches the uniform distribution of unitaries up to the second moment (i.e., forms a 2-design), then the variance in the cost gradient will vanish exponentially with the number of qubits. Several strategies have been proposed to address this issue~\cite{Volkoff2020BP,Grant2019initializationBP,verdon2019learning,skolik2020layerwise,pesah2020absence,zhang2020toward,campos2020abrupt,patti2020entanglement,bharti2020iterative,bharti2020quantumAS}, such as clever parameter initialization or ansatz construction, while more research is needed to test these strategies on various problems.

In broad terms, the expressibility of an ansatz is determined by how uniformly it explores the unitary space. 
Thus the distance between the distribution of unitaries generated by an ansatz and the maximally expressive uniform distribution of unitaries is a natural measure of its expressibility~\cite{ExpressibilitySukin2019}.
Using such a measure, Ref.~\cite{ExpressibilityNakaji2020} calculated the expressiblity for several commonly used ans\"{a}tze and, by using the cost gradients obtained in \cite{cerezo2020cost}, suggested that in some cases it is possible for an ansatz to be both expressive and trainable. Additionally, Ref.~\cite{ExpressibilityTangpanitanon2020} noted a numerical correlation between expressibility and trainability for analog systems. However, given that both expressibility and trainability are closely related to randomness, one might expect to be able to draw a more fundamental and general relationship between expressibility and trainability.

Here we demonstrate that this is indeed the case by analytically relating the trainability of an ansatz to its expressibility. This is done by extending the barren plateau phenomenon introduced in \cite{mcclean2018barren}, which holds for ans\"{a}tze that form exact 2-designs, to arbitrary ans\"{a}tze. 
Specifically, we upper bound the variance in the cost gradient in terms of the distance the ansatz is from being a 2-design. Since the degree to which an ansatz is a 2-design is a measure of its expressibility, this allows us to relate the gradient of the cost landscape to the expressibility of the ansatz. We find that the more expressive the ansatz, the smaller the variance in the cost gradient and hence the flatter the landscape. We note that an ansatz does not strictly need to be highly expressive to be used successfully, rather it just needs to contain a solution to the problem at hand. Thus our result highlights the importance of developing trainable problem-inspired ansatze.

\begin{figure}[t!]
    \centering
    \includegraphics[width=0.95\columnwidth]{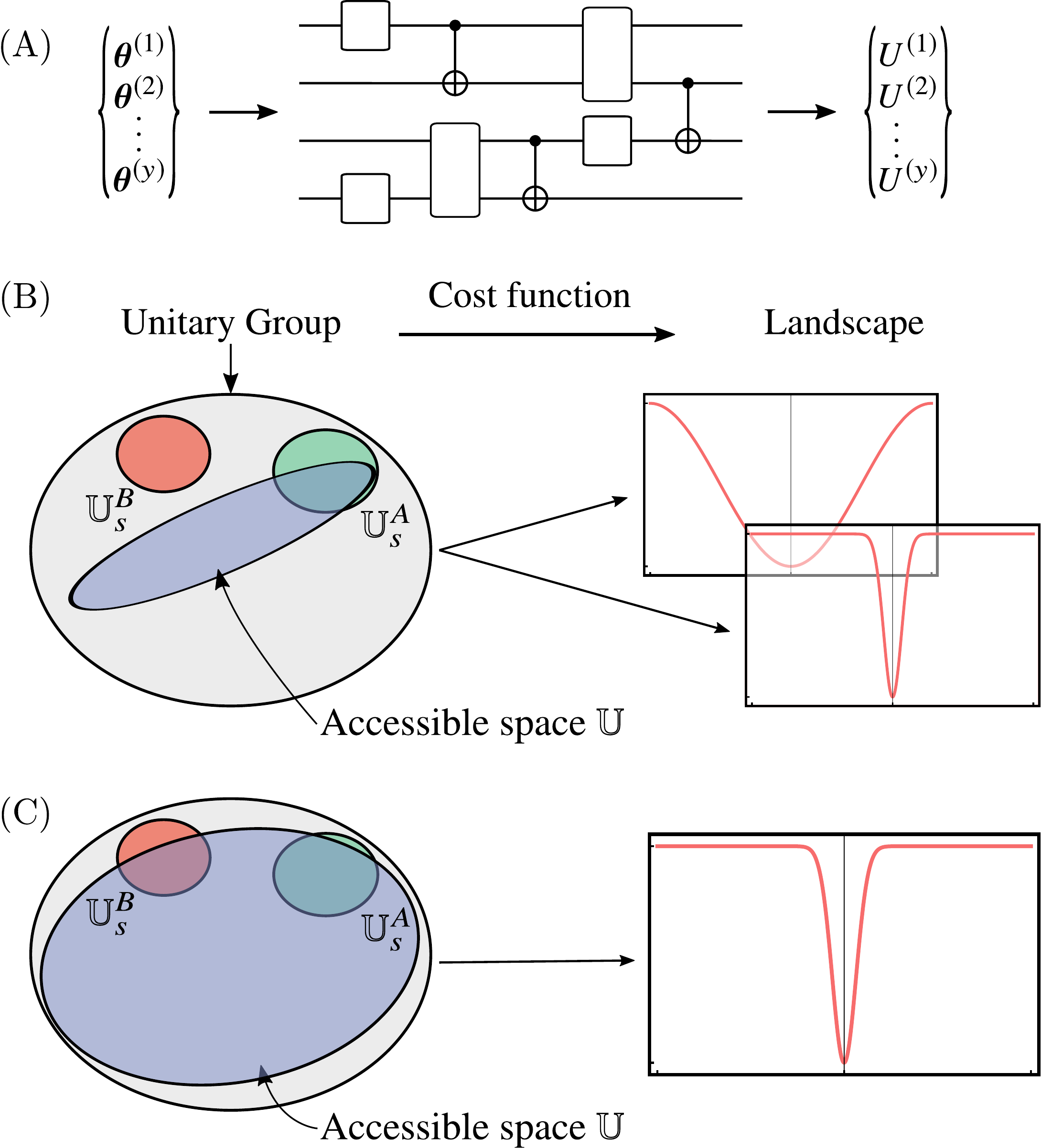}
    \caption{\textbf{Schematic representation of  main results.} (A) Variational quantum algorithms (VQAs) train the parameters $\thv$ in a parameterized quantum circuit  to minimize a cost function as in Eq.~\eqref{eq:GenCost}. Each set of parameters corresponds to a unitary $U(\thv)$ being produced. The set of unitaries $\mathbb{U}$ accessible by $U(\thv)$ is a subset of the unitary group $\UC(d)$, and the VQA can be successful if $\mathbb{U}$ overlaps with the space of solution unitaries $\mathbb{U}_s$ that (approximately) minimize the cost. The expressibility of an ansatz quantifies the degree to which it uniformly explores the unitary group $\UC(d)$. Given problems $A$ and $B$, we denote their solution spaces as $\mathbb{U}_s^A$ and $\mathbb{U}_s^B$ respectively. (B) A low-expressibility ansatz contains solutions to problem $A$ but not to $B$, while a high-expressibility ansatz as in (C) contains solutions to both problems. Low-expressibility ans\"{a}tze can lead to both small and large cost gradients. On the other hand, high-expressibility ans\"{a}tze lead to predominantly flat cost landscapes, and thus are generally hard to train.
    }\label{fig:Schematic}
\end{figure}

Our main results can be summarized in Fig.~\ref{fig:Schematic}. Given an ansatz, we analyze the space of unitaries accessible  when  sampling the parameters (Fig.~\ref{fig:Schematic}(A)) of a parametrized quantum circuit. Inexpressive ans\"{a}tze, such as the one shown in Fig.~\ref{fig:Schematic}(B), access a small region of the unitary group and can include the space of unitaries that solve certain problems but not the space that solve others.  Our results do not preclude inexpressive ans\"{a}tze having trainability issues, such as barren plateaus. On the other hand, highly expressive ans\"{a}tze, which are generically used for many problems as they can access a much larger space (Fig.~\ref{fig:Schematic}(C)), are shown to lead to small gradients, and hence can have trainability issues.

Since our analytic bounds are upper bounds, they leave open the questions of how reducing the expressiblity of an ansatz changes the cost landscape, and hence how reducing the expressibility can be used to avoid the barren plateau phenomenon. To address these questions we provide extensive numerics studying the effect that tuning the expressibility of an ansatz may have on the scaling of gradient magnitudes. Specifically, we consider the effects of decreasing the depth of the circuits, correlating circuit parameters, and restricting either the direction or angle of rotations. We find that strongly correlating parameters~\cite{Volkoff2020BP} and/or initializing close to the solution (and then restricting the ansatz to explore the region close to the initialization~\cite{Grant2019initializationBP}) to be the most effective approaches to avoid exponentially vanishing cost gradients.  

\section{Preliminaries}

\subsection{General framework} Variational Quantum Algorithms (VQAs) encode an optimization task in a cost function whose minimum corresponds to the solution of the problem.  Here we consider cost functions of the form\footnote{In Appendix~\ref{ap:Derivations} we extend our results to more general costs of the form $C_{\rm gen}   = \sum_i \Tr[ H_i U(\vec{\theta})  \rho_i U(\vec{\theta}) ^\dagger ] $. This cost allows for multiple input states $\{\rho_i \}$ and measurement operators $\{ H_i \}$, opening up quantum machine learning approaches that employ training data~\cite{QMLBiamonte2017,schuld2015introduction,QNFLPoland2020, SharmaQNFL2020}.}  
\begin{equation}\label{eq:GenCost}
   C_{\rho, H}( \vec{\theta})   =  \Tr[ H U(\vec{\theta})  \rho U(\vec{\theta}) ^\dagger ] \, ,
\end{equation}
where $\rho$ is an $n$-qubit input state, $H$ is a Hermitian operator, and $U(\thv)$ is a parametrized quantum circuit depending on trainable parameters $\vec{\theta}$. The value of the cost $C_{\rho, H}( \vec{\theta}) $ (or of its gradient) are estimated on a quantum computer, and then are fed into a classical optimizer which attempts to solve the optimization task $\argmin_{\vec{\theta}} C_{\rho, H}( \vec{\theta})$.

The success of the VQA hinges on several factors. First, it is necessary to find an operator $H$ such that the resulting cost is faithful for the given problem. That is, we require the minimum of $C_{\rho, H}( \vec{\theta})$ to correspond to the solution of the optimization task. Evidently, for some applications, there may be multiple choices in $H$ corresponding to faithful costs and therefore other factors will determine which to use. One such factor is how easily $H$ can be measured on a quantum computer. Another relevant feature, as discussed further in Section~\ref{sec:PrelimBPs}, is the {\it locality} of $H$, i.e., the number of qubits it acts non-trivially on. We say that the cost function is {\it global} if $H$ acts non-trivially on all qubits, while we use the term {\it $k$-local} for costs where $H$ acts non-trivially on at most $k$ qubits.

A second aspect that determines the success of a VQA is the choice in {\it ansatz} for $U(\vec{\theta})$. 
While discrete parameterizations are possible, usually $\vec{\theta}$ are continuous parameters, such as gate rotation angles, in a parametrized quantum circuit.
Generally,  $U(\thv)$ is expressed as 
\begin{equation}\label{eq:LayeredAnsatz}
    U(\vec{\theta}) = \prod_{j=1}^D U_j(\theta_j) W_j .
\end{equation}
Here $\{W_j \}_{j=1}^N$ is a chosen set of fixed unitaries and $U_j = e^{-i \theta_j V_j}$ is a rotation of angle $\theta_j$ generated by a Hermitian operator $V_j$ such that $(V_j)^2 = \I$. The rotation angles $\{ \theta_j \}$ are typically assumed to be independent.  

Once an ansatz has been fixed for the parametrized quantum circuit, then, as sketched in Fig.~\ref{fig:Schematic}(A), each possible vector of parameters $\vec{\theta}$ corresponds to a unitary $U(\vec{\theta})$ that is produced. For concreteness, given a set of different parameters $\{ \thv^{(1)}, ... \thv^{(2)}, ... , \thv^{(y)} \}$ we obtain the corresponding ensemble of unitaries 
\begin{align}\label{eq:ensembleU}
\mathbb{U} = \{ U^{(1)}, U^{(2)}, ... , U^{(y)} \},
\end{align} 
where $ U^{(j)} := U(\thv^{(j)})$. Here, $\mathbb{U}\subseteq \UC(d)$, where $\UC(d)$ is the unitary group $\UC(d)$ of degree $d=2^n$.

\medskip

\subsection{Expressibility} 

For a VQA to be successful, a solution (i.e., a unitary which is by some measure close to the unitary that minimizes the cost) needs to be contained within the ensemble of unitaries generated by the ansatz. Specifically, defining $\mathbb{U}_s$ as the set of solution unitaries, then the VQA will be successful only if $\mathbb{U}_s \bigcap \mathbb{U}\neq \emptyset$. When this condition is satisfied the ansatz is said to be \textit{complete} for the given problem. 

In the absence of prior knowledge about where the solution unitaries $\mathbb{U}_{s}$ lie, the likelihood that the ansatz is complete can be maximized by using an ansatz that explores the total space of unitaries as fully and as uniformally as possible. Such ans\"{a}tze are known as \textit{expressive} ans\"{a}tze. For example, consider having two problems (problem $A$, and problem $B$), with solution spaces respectively denoted as $\mathbb{U}_s^A$ and $\mathbb{U}_s^B$. Figure~\ref{fig:Schematic}(B) sketches $\mathbb{U}$ for an inexpressive ansatz which is complete with respect to problem $A$ but incomplete with respect to $B$. Conversely, Fig.~\ref{fig:Schematic}(C) shows $\mathbb{U}$ for an expressive ansatz which is complete with respect to both problems. 

For many applications, information about the problem can be encoded in the ansatz.
For instance, the quantum alternating operator ansatz~\cite{hadfield2019quantum} (or the Hamiltonian variational ansatz~\cite{wecker2015progress}), encode information of an appropriate adiabatic transformation. 
Such \textit{problem-inspired} ans\"{a}tze may be complete but inexpressive (e.g., Fig.~\ref{fig:Schematic}(B) could denote a problem-inspired ansatz for problem $A$). However, \textit{problem-agnostic} ans\"{a}tze, which can be used for a wide range of problems, need to be sufficiently expressive to guarantee their completeness. 

The expressibility of an ansatz, i.e., the degree to which it uniformly explores the unitary group $\UC(d)$, can be quantified by comparing the uniform distribution of unitaries obtained from the ensemble $\mathbb{U}$ to the maximally expressive uniform (Haar) distribution of unitaries from  $\mathcal{U}(d)$. 
More concretely, the expressibility of a circuit can be defined in terms of  the following super-operator~\cite{ExpressibilitySukin2019,ExpressibilityNakaji2020}:
\begin{multline}\label{eq:expressibility}
    \mathcal{A}^{(t)}_{\mathbb{U}}(\cdot)
    := \int_{\mathcal{U}(d)} d\mu(V)  V^{\otimes t}  ( \,\cdot \, ) (V^\dagger)^{\otimes t}\\  -  \int_{\mathbb{U}} d U \, U^{\otimes t} ( \,\cdot\, ) (U^\dagger)^{\otimes t} \, ,
\end{multline}
where $d\mu(V)$ is the volume element of the Haar measure and $dU$ is the volume element corresponding to the uniform distribution over $\mathbb{U}$ in Eq.~\eqref{eq:ensembleU}. 
If $ \mathcal{A}^{(t)}_{\mathbb{U}}(X) =  0$ for all operators $X$, then averaging over elements of $\mathbb{U}$ agrees with averaging over elements of the Haar distribution over $\mathcal{U}(d)$ up to the $t$-th moment, and thus $\mathbb{U}$ forms a $t$-design~\cite{tDesign2002, tDesign2007, ChaosByDesign, LowThesis, 2019HunterJones}.
For our purposes it suffices to consider the behavior of $\mathcal{A}^{(t)}_{\mathbb{U}}$ for $t=2$. Henceforth we drop the t-superscript, i.e., $\mathcal{A}_{\mathbb{U}} \equiv \mathcal{A}_{\mathbb{U}}^{(2)}$. 

In the context of minimizing a generic cost $C_{\rho, H}( \vec{\theta})$ of the form specified by Eq.~\eqref{eq:GenCost}, we are interested in the expressibility of the circuit with respect to both the initial state $\rho$ and the measurement operator $H$. The following quantities respectively capture these notions:
\begin{align}
    &\varepsilon_{\mathbb{U}}^\rho := || \AC_{{\mathbb{U}}}( \rho^{\otimes 2}) ||_2 \\
    &\varepsilon_{\mathbb{U}}^H := || \AC_{{\mathbb{U}}}( H^{\otimes 2}) ||_2 \, .
\end{align}
Small values of $\varepsilon_{\mathbb{U}}^\rho$ and $\varepsilon_{\mathbb{U}}^H$ indicate that the ansatz is highly expressive. 
These measures generalize the notion of expressibility introduced in \cite{ExpressibilitySukin2019} where the expressiblity was defined in terms of 
$\varepsilon_{\mathbb{U}}^{\rho}$ for $\rho = |0\rangle \langle 0|$.

While the $\rho$ and $H$ dependence of $\varepsilon_{\mathbb{U}}^\rho$ and $\varepsilon_{\mathbb{U}}^H$ make them natural measures of the expressibility in the context of minimizing a cost $C_{\rho, H}( \vec{\theta})$, cost function-independent measures of expressibility may allow the expected performance of different ans\"{a}tze to be more easily compared.
With this in mind, one could alternatively quantify the expressiblity directly in terms of the diamond norm of $\AC_{\mathbb{U}}$,
\begin{equation}\label{eq:eps-diamond}
    \varepsilon^{\scriptscriptstyle \diamond}_{\mathbb{U}} := || \mathcal{A}_{\mathbb{U}} ||_{\diamond} \,  ,
\end{equation}
which is an operationally meaningful distance measure to distinguish two quantum operations. We use the diamond norm here in line with the literature on $\varepsilon$-approximate unitary designs~\cite{HarrowLowApproxDesigns09}; however, alternative norms can be used (for a discussion, see \cite{LowThesis}). 
For completeness we will formulate our results in terms of $ \varepsilon_{\mathbb{U}}^{\scriptscriptstyle \diamond}$, as well as the quantities $\varepsilon_{\mathbb{U}}^\rho$ and $\varepsilon_{\mathbb{U}}^H$.

\medskip

\subsection{Gradient Magnitudes} For a variational quantum algorithm to run successfully it is not sufficient that the ansatz contains the solution; the cost landscape must also exhibit large enough cost gradients to enable this solution to be found. 

The component of the gradient corresponding to the parameter $\theta_k$ is determined by the partial derivative $\partial_k C := \frac{\partial C_{\rho, H}( \vec{\theta})}{\partial \theta_k }$. For a generic ansatz of the form specified by Eq.~\eqref{eq:LayeredAnsatz}, the average of $\partial_{k} C$ over all parameters $\thv$ vanishes
\begin{equation}\label{eq:GradUnbiased}
\begin{aligned}
      \langle  \partial_{k} C \rangle &=0  \ \ \ \forall \ \ k \, .
\end{aligned}
\end{equation}
That is, the cost gradients are not biased in any single direction but rather average out to zero.
Intuitively, this lack of bias can be understood as following from the fact that the average of a rotation $\exp(-i \theta_k V_k)$ is zero when $V_k^2 = \id$. 
We show this in Appendix~\ref{ap:Unbiased}, where we prove that $\langle  \partial_{k} C \rangle =0 $ by explicitly integrating over $\theta_k$. 

However an unbiased cost landscape can be either trainable or untrainable, depending on the extent to which the gradient fluctuates away from zero. Therefore, to assess the trainability of an ansatz $U(\thv)$, we now recall the Chebyshev inequality. This inequality bounds the probability that the partial derivative of the cost deviates from its average of zero, 
\begin{equation}\label{eq:Cheb}
    P( |\partial_{k} C | \geq \delta ) \leq \frac{\text{Var}[\partial_{k} C]}{\delta^2} \, ,
\end{equation}
in terms of the variance 
\begin{equation}
    \Var[\partial_{k} C] = \left\langle  \left(\partial_{k} C\right)^2 \right\rangle - \left\langle  \partial_{k} C \right\rangle^2  \,, 
\end{equation}
where the expectation value is taken over the parameters $\thv$. Hence if the variance of the partial derivative is small for all $\theta_k$, then the probability that the partial derivative is non-zero is small for all $\theta_k$. On such landscapes, (potentially untenably) precise measurements are required to detect the path of steepest descent to navigate to the minimum. 

\medskip

\subsection{Barren Plateaus}\label{sec:PrelimBPs}

There is a growing awareness of the so called barren plateau phenomenon for variational quantum algorithms~\cite{mcclean2018barren,cerezo2020cost,sharma2020trainability,wang2020noise,cerezo2020impact,holmes2020barren,marrero2020entanglement,uvarov2020barren,arrasmith2020effect,abbas2020power}. For a given ansatz $U(\thv)$, a cost $C$ is said to exhibit a barren plateau if its gradients vanish exponentially with the number of qubits $n$. This is typically relaxed to a probabilistic definition, where the gradient vanishes exponentially with high probability. This would follow from Chebyshevs inequality, Eq.~\eqref{eq:Cheb}, if the variance in the partial derivative vanishes exponentially, i.e., if $\Var[\partial_{k} C] \in \mathcal{O}(2^{- p n})$ for any integer $p > 0$. For costs that exhibit barren plateaus, exponentially precise measurements may be required to determine the minimization direction, and hence the cost is effectively untrainable for large problem sizes. 

To elucidate the conditions under which a layered parameterized ansatz $U(\thv)$, of the form of Eq.~\eqref{eq:LayeredAnsatz}, gives rise to barren plateaus, consider a bipartite cut of $U(\thv)$ and write 
\begin{equation}\label{eq:structure}
    U(\vec{\theta}) = U_{L}(\vec{\theta}) U_R(\vec{\theta}) 
\end{equation}
where 
\begin{align}\label{eq:ulur}
 U_L(\vec{\theta}) = \prod_{j=k+1}^D U_j(\vec{\theta_j}) W_j \ \ \ \text{and} \ \ \ U_R(\vec{\theta}) = \prod_{j=1}^k U_j(\vec{\theta}_j) W_j \, .
\end{align}
Note that since we suppose the parameters $\theta_j$ are uncorrelated, the circuits $U_L$ and $U_R$ are independent. These circuits are pertinent when quantifying gradients since taking the partial derivative of a circuit, as shown in Appendix~\ref{ap:Derivations},  effectively splits a circuit in two.

Ref.~\cite{mcclean2018barren} then demonstrated that if the ensemble of unitaries generated by the ansatz $U(\thv)$ is sufficiently random (i.e., expressive) such that the ensembles $\mathbb{U}_L$ or $\mathbb{U}_R$ (associated with the circuits $U_L(\thv)$ and $U_R(\thv)$ respectively) form 2-designs, then the variance in the cost gradient vanishes exponentially with $n$. Specifically,
let us denote the variance of the cost when just $\mathbb{U}_R$, just $\mathbb{U}_L$, and both $\mathbb{U}_R$ and $\mathbb{U}_L$ form 2-designs as $\text{Var}_{R}\partial_k C $, $\text{Var}_{L}\partial_k C $, and $\text{Var}_{R, L}\partial_k C$, respectively. From Ref.~\cite{mcclean2018barren} it follows that for $x = R$, $x = L$ and $x = R, L$,
\begin{align}
         & \text{Var}_x \partial_k C = \frac{g_x(\rho, H, U)}{2^{2n}-1} ~,  \label{eq:GoogleVariance}
\end{align}
where we have pulled out the $n$-dependent scaling factor explicitly. The prefactor $g_x(\rho, H, U)$, which we define explicitly in Appendix~\ref{sec:google-high-order}, is in $\OC(2^n)$ for typical choices in $V_k$ and $H$.  
Thus if $\mathbb{U}_L$ or $\mathbb{U}_R$ form a 2-design, the variance in the gradient vanishes exponentially in $n$. In other words, maximally expressive ans\"{a}tze exhibit barren plateaus.

\section{Main Results}

\subsection{Analytic Bounds}

In this section, we study the gradient of a generic cost $C_{\rho, H}( \vec{\theta})$, Eq.~\eqref{eq:GenCost}, with an ansatz $U(\thv)$, Eq.~\eqref{eq:LayeredAnsatz}, but relax the assumption that $\mathbb{U}_L$ or $\mathbb{U}_R$ forms a 2-design. By doing so, we extend the results on barren plateaus from Ref.~\cite{mcclean2018barren} to arbitrary ans\"{a}tze. As will become clear, this generalization enables us to relate the variance of the cost function partial derivative to the expressiblity of $U(\thv)$ in Eq.~\eqref{eq:expressibility}.

Let us start by noting that while maximally expressive ans\"{a}tze exhibit barren plateaus, the converse is not necessarily true. In other words, highly inexpressive ans\"{a}tze need not always experience large cost gradients, and in fact they may exhibit vanishing gradients. A trivial example of this phenomenon is provided by an ansatz composed of rotations that commute with the measurement operator $[U(\theta), H] = 0$. Such an ansatz will leave the cost unchanged for any $\theta$ and so the variance in gradient in the cost of such an ansatz is necessarily zero. A more subtle example is an ansatz composed of a tensor product of single qubit rotations. Since this ansatz does not generate entanglement it is inexpressive; however, it has also been shown to exhibit a barren plateau for global cost functions~\cite{khatri2019quantum, cerezo2020cost}. It follows from these observations that it is not possible to meaningfully \textit{lower} bound the gradients of an ansatz in terms of its expressiblity. 

Therefore to relate cost gradients to expressibility we instead derive an \textit{upper} bound. Specifically, our main result consists of a non-trivial upper bound for the variance of the cost function partial derivative for a general ansatz $U(\thv)$ in terms of the expressibility in~\eqref{eq:expressibility}.
This bound is in terms of: (1) the variance of the cost gradient when either $\mathbb{U}_L$ or $\mathbb{U}_R$ form a 2-design, and (2) the expressibility of the ansatz as measured by the distance $\mathbb{U}_L$ and $\mathbb{U}_R$ are from being 2-designs. As shown in Appendix~\ref{ap:Derivations}, we prove the following.
\begin{thm}\label{thm:mainbounds}
Consider a generic cost function $C_{\rho, H}( \vec{\theta})$, Eq.~\eqref{eq:GenCost}, using a layered ansatz $U(\vec{\theta})$ of the general form in Eq.~\eqref{eq:LayeredAnsatz}. The variance of the cost partial derivative obeys the following bounds: 
\begin{align}
   &  \Var \, \partial_k C  \leq \Var_R \, \partial_k C  +  4  \varepsilon_R^\rho ||H||_{2}^2~, \label{eq:MainBoundR} \\
   &  \Var \, \partial_k C   \leq   \Var_L \, \partial_k C + 4\varepsilon_L^H  ||\rho||_{2}^2~,  \label{eq:MainBoundL} \\
   &  \Var \, \partial_k C  \leq   \Var_{R,L} \, \partial_k C   +  f( \varepsilon_R^\rho , \varepsilon_L^H) \label{eq:MainBoundLR}  \, .
\end{align}
Here we used the shorthand $\varepsilon_R^\rho := \varepsilon_{{\mathbb{U}}_R}^\rho$ and $\varepsilon_L^H :=  \varepsilon_{{\mathbb{U}}_L}^H$, and we have defined 
\begin{equation}\label{eq:f(x,y)}
    f(x, y) := 4 x y  + \frac{2^{n+2} \left( x ||H||_{2}^2 +  y ||\rho||_{2}^2 \right)}{2^{2n} - 1} \, .
\end{equation}.
\end{thm}

Theorem~\ref{thm:mainbounds} establishes a formal relationship between the gradient of the cost landscape and the expressibility of the ansatz used. Namely, the higher the expressibility of the ansatz, that is the smaller $\varepsilon_L^H$ or $\varepsilon_R^\rho$, the smaller the upper bound on the variance of the cost partial derivative. This, in combination with the fact that the cost gradient is unbiased, demonstrates that highly expressive ans\"{a}tze will have flatter landscapes and consequently be harder to train.

In contrast to the bounds specified by Eqs.~\eqref{eq:GoogleVariance}, which hold for three distinct cases (i.e., when $\mathbb{U}_L$ is a 2-design, when $\mathbb{U}_R$ is a 2-design, and when both $\mathbb{U}_L$ and $\mathbb{U}_R$ are 2-design), the bounds in Eqs.~\eqref{eq:MainBoundR}--\eqref{eq:MainBoundLR} all hold for any generic ansatz of the form in Eq.~\eqref{eq:LayeredAnsatz}. Thus any single bound would suffice to bound the variance in the cost function partial derivative for an arbitrary ansatz. 

We include all three bounds despite this fact since in any instance one bound may be tighter than the others and hence more informative.
In particular, the relative tightness of the bounds depends on which parameter we are taking the derivative with respect to. This follows from the fact that Eq.~\eqref{eq:MainBoundR} becomes an equality in the limit that $\mathbb{U}_R$ tends to a 2-design, where as Eq.~\eqref{eq:MainBoundL} becomes an equality in the limit that $\mathbb{U}_L$ is a 2-design and Eq.~\eqref{eq:MainBoundLR} becomes an equality in the limit that both $\mathbb{U}_L$ and $\mathbb{U}_R$ are 2-designs. If we are looking at the derivative with respect to the final layer then $\mathbb{U}_R$ is typically closer to being a 2-design than $\mathbb{U}_L$ and so \eqref{eq:MainBoundR} will be tightest. Conversely, if we are most interested in the partial derivative with respect to a parameter in the first layer then \eqref{eq:MainBoundL} will be tightest. On the other hand, for parameters in a layer close to the middle (i.e. at depth $D/2$) and \eqref{eq:MainBoundLR} will be tightest since, as shown in Appendix~\ref{ap:Derivations}, the derivation of this bound uses the most information about the ansatz.

In Appendix~\ref{ap:Derivations}, we extend Theorem~\ref{thm:mainbounds} to cost functions of the form $C_{\rm gen}   = \sum_i \Tr[ H_i U(\vec{\theta})  \rho_i U(\vec{\theta})^\dagger]$, which allow for multiple input states and measurements. Thus our results also apply to quantum machine learning approaches that utilize training data~\cite{QMLBiamonte2017,schuld2015introduction,QNFLPoland2020, SharmaQNFL2020}.  

\medskip

\paragraph*{Generalizing the Barren Plateau phenomenon.}
Theorem~\ref{thm:mainbounds} may be viewed as an extension of the barren plateau phenomenon introduced in Ref.~\cite{mcclean2018barren} to ans\"{a}tze that form approximate, rather, than exact 2-designs. 
By combining Eq.~\eqref{eq:GoogleVariance} and Eq.~\eqref{eq:MainBoundLR}, 
we find that the variance in the partial derivative for an arbitrary ansatz is bounded as
\begin{align}
 \Var \, \partial_k C  \leq  \frac{g_{L,R}(\rho, H, U)}{2^{2n}-1} +  f(\varepsilon_L^H , \varepsilon_R^\rho) \label{eq:MainBoundLRcorol}  \, .
\end{align}
Here the first term on the right is the variance of a maximally expressive ansatz (namely, one that forms a 2-design) and $f(\varepsilon_L^H , \varepsilon_R^\rho)$ is the expressiblity dependent correction term defined in Eq.~\eqref{eq:f(x,y)}. Expressions similar to Eq.~\eqref{eq:MainBoundLRcorol} are obtainable from Eq.~\eqref{eq:MainBoundR} and Eq.~\eqref{eq:MainBoundL}. 

For perfectly expressive ans\"{a}tze, $f(\varepsilon_L^H , \varepsilon_R^\rho)$ vanishes and Eq.~\eqref{eq:MainBoundLRcorol} reduces to Eq.~\eqref{eq:GoogleVariance}, regaining the result of Ref.~\cite{mcclean2018barren}. In this case, the variance in the gradient vanishes exponentially with the size of the system $n$, i.e., the ansatz exhibits a barren plateau. Similarly, if the expressibility of an ansatz increases exponentially with the size of the problem, i.e., if $ f(\varepsilon_L^H , \varepsilon_R^\rho) \in \OC \left(\frac{1}{2^{k n}}\right)$ for $k > 0$, then $ \Var \, \partial_k C $ again vanishes exponentially and the ansatz exhibits a barren plateau. However, more generally, when $ f(\varepsilon_L^H , \varepsilon_R^\rho)$ scales non-exponentially the upper bound allows for the variance in the partial derivative to be non-vanishing. Thus, there is leeway for imperfectly expressive ans\"{a}tze to avoid barren plateaus. 

In Ref.~\cite{arrasmith2021equivalence} it was proven that the barren plateau phenomenon is necessarily associated with the concentration of cost functions values about their mean. More concretely, it was shown that the probability that the cost function deviates from its mean is determined by the variation in the gradient of the cost. Thus our bounds also imply that the degree to which the cost concentrates about its mean increases with increasing expressibility. In Appendix~\ref{ap:Conc}, we provide an alternative proof of this following on from the results of Ref.~\cite{marrero2020entanglement}.

\medskip

\paragraph*{Diamond Norm Reformulation.}
For local costs the term $||H||_2^2$ scales exponentially with the size of the system and therefore for large systems \eqref{eq:MainBoundR} becomes exponentially loose. This issue can be mitigated by reformulating Theorem~\ref{thm:mainbounds} in terms of $\varepsilon_{\mathbb{U}}^{\diamond}$, Eq.~\eqref{eq:eps-diamond}. We obtain the following theorem in Appendix~\ref{ap:Derivations}. 

\begin{thm}\label{corol:DiamondNormBounds}
Consider a generic cost function $C_{\rho, H}( \vec{\theta})$, Eq.~\eqref{eq:GenCost}, using a layered ansatz $U(\vec{\theta})$ of the general form in Eq.~\eqref{eq:LayeredAnsatz}. The variance of the cost partial derivative obeys the following bounds: 
\begin{align}
\Var \, \partial_k C & \leq \Var_R \, \partial_k C  + 4 || H ||_\infty^2 \, \eRd ~, \label{eq:boundRd} \\
\Var \, \partial_k C & \leq \Var_L \, \partial_k C  + 4|| \rho ||_\infty^2 \Vert H \Vert_1 \, \eLd ~, \label{eq:boundLd} \\
\Var \, \partial_k C  &\leq  \Var_{R,L} \, \partial_k C  + \frac{ f(\eRd, \Vert H\Vert_1\eLd)}{2^{n}} \label{eq:MainBoundLRd} \, ,
\end{align}
where we use the shorthand $\eRd = \varepsilon_{{\mathbb{U}}_R}^{\diamond}$ and $\eLd = \varepsilon_{{\mathbb{U}}_L}^{\diamond}$ and with $f(x,y)$ defined in Eq.~\eqref{eq:f(x,y)}.
\end{thm}

Again, Theorem~\ref{corol:DiamondNormBounds} formally establishes that highly expressive ans\"{a}tze experience flatter cost landscapes. Furthermore, a relation similar to \eqref{eq:MainBoundLRcorol} can be derived from Theorem~\ref{corol:DiamondNormBounds}. Hence, Theorem~\ref{corol:DiamondNormBounds} also provides an extension of the barren plateau result of Ref.~\cite{mcclean2018barren}. However, since $|| H ||_\infty^2 \in \mathcal{O}(1)$ for all $H$,  \eqref{eq:boundRd} does not experience the same looseness for local costs of large systems as \eqref{eq:MainBoundR}. On the other hand, since $||H||_1$ may scale exponentially in $n$, \eqref{eq:boundLd} may become loose for large systems and therefore we expect \eqref{eq:MainBoundL} to generally be more useful than \eqref{eq:boundLd}.

\begin{figure}[t!]
    \centering
    \includegraphics[width=0.48\textwidth]{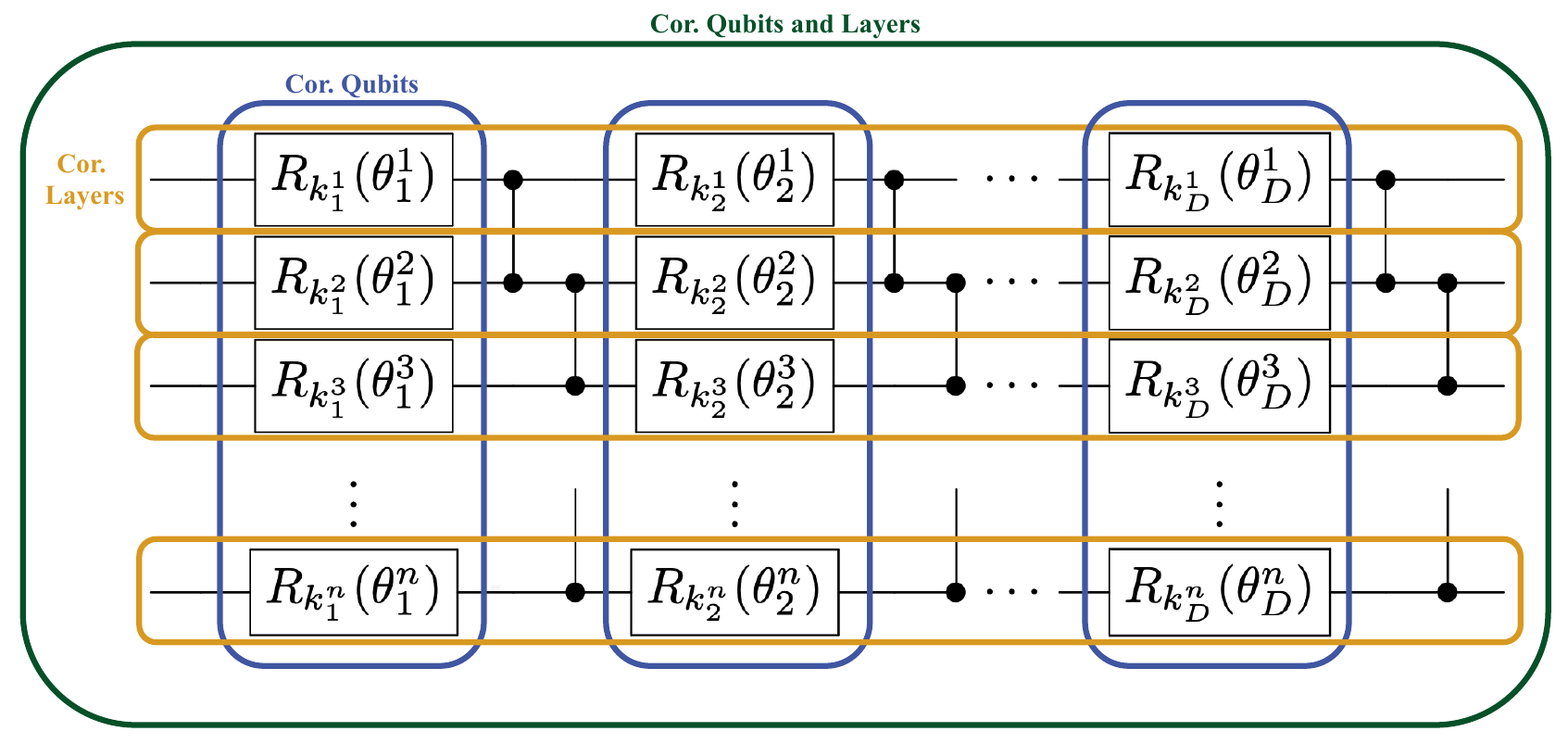}
    \caption{\textbf{Ansatz employed in numerical simulations}. The ansatz is composed of alternating random single qubit rotations and ladders of $\text{C-Phase}$ operations. The colored boxes indicate the gates which are fixed to rotate by the same angle and in the same direction when we correlate the ansatz layers (yellow), correlate qubits (blue) and correlate both the layers and qubits (green). 
    }\label{fig:Ansatz}
\end{figure}

\subsection{Numerical Simulations}\label{sec:Numerics}

Since the analytic bounds in the previous section are upper bounds, we have no guarantee that inexpressive ans\"{a}tze will exhibit larger cost gradients. The bounds thus leave open the question of whether/how reducing the expressiblity of an ansatz changes the cost landscape. Moreover, they leave open the question of how one can avoid the barren plateau phenomenon that is observed for maximally expressive ans\"{a}tzes.

\begin{figure*}[t!]
    \centering
    \includegraphics[width=\textwidth]{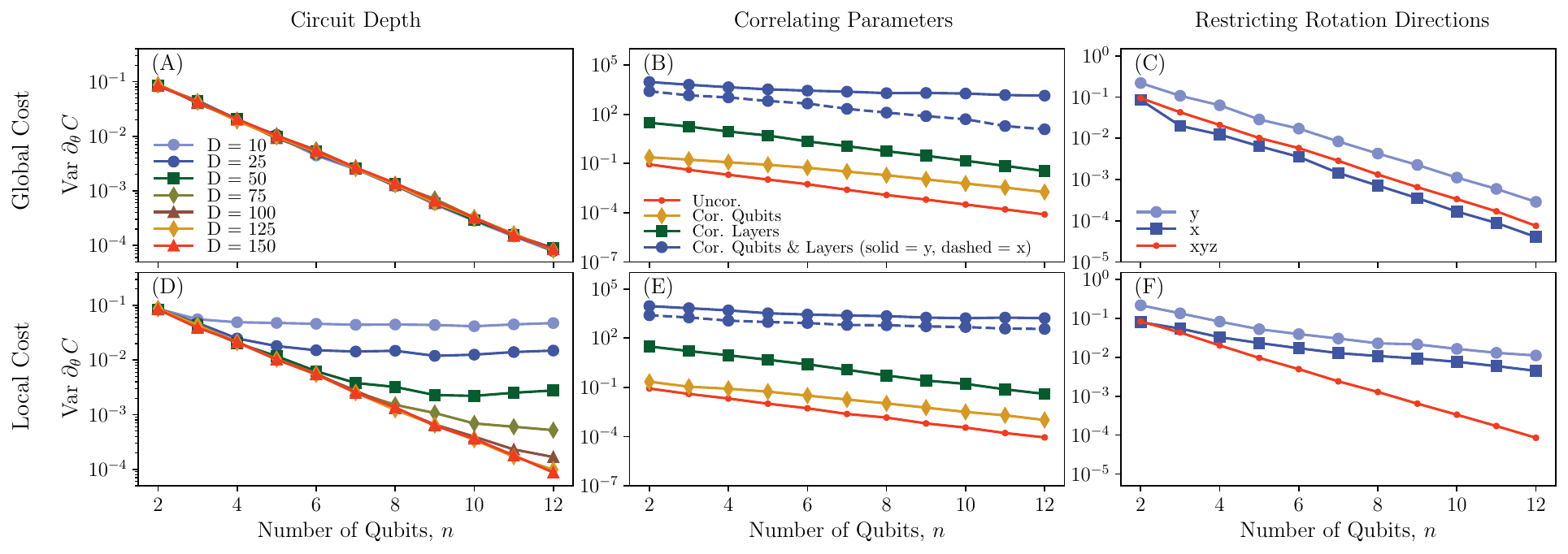}
    \caption{\textbf{Partial derivative scalings for different expressiblities.} The variance in the partial derivative of a global cost with $H_G = \bigotimes_{i=1}^n \sigma^z_i$ (top) and 2-local cost with $H_L = \sigma^z_1 \sigma^z_2$ (bottom) as a function of the number of qubits $n$ (in both cases $\rho = |\psi_0\rangle \langle \psi_0 |^{\otimes n}$ where $\ket{\psi_0} = \exp(-i(\pi/8)\sigma_Y) \ket{0}$). In the left panel we vary the circuit depth $D$ of a hardware efficient ansatz. In the middle (right)} panel we consider the effect of correlating parameters (restricting the directions of rotation) of a hardware efficient ansatz with $D =150$ with the choices of correlations (rotations) indicated in the figure legend. In all cases the derivative is taken with respect to $\theta_1^1$, the rotation angle of the first qubit in the first layer, and the variance is taken over an ensemble of $1000$ unitaries.\label{fig:VarGradPanels}
\end{figure*}

One can conceive of numerous ways in which the expressibility of an ansatz can be tuned, each of which could have a different impact. In this section, we consider four such ways: decreasing the depth of the circuits, correlating circuit parameters, and restricting either the direction or angle of rotations. We then numerically investigate the effect these have on the cost gradient scaling.

For completeness, in our numerics we consider both a 2-local cost where the measurement operator is composed of Pauli-$z$ measurements on the first and second qubits, $H_L = \sigma^z_1 \sigma^z_2$, and a global cost where the measurement operator consists of Pauli-$z$ measurements across all qubits, $H_G = \bigotimes_{i=1}^n \sigma^z_i$~\cite{cerezo2020cost}. 
In both cases, following \cite{mcclean2018barren}, the system is prepared in the pure state, $\rho = |\psi_0\rangle \langle \psi_0 |^{\otimes n} $ where $\ket{\psi_0} = \exp(-i(\pi/8)\sigma_Y) \ket{0}$. 
We further consider a layered hardware efficient ansatz,
\begin{equation} \label{eq:HEA}
    U(\vec{k}_l, \thv_l, D) := \prod_{l=1}^D W V(\vec{k}_l, \thv_l) \, ,
\end{equation}
consisting of $D$ alternating layers of random single qubit gates and entangling gates as shown in Fig.~\ref{fig:Ansatz}.  
Specifically, the entangling layer,
\begin{equation}
    W = \prod_{i = 1}^{n-1} \text{C-Phase}_{i, i+1} \, ,
\end{equation}
is composed of a ladder of controlled-phase operations, $\text{C-Phase}$, between adjacent qubits in a 1-dimensional array.
The single-qubit layer consists of a series of random single qubit rotations
\begin{equation}
    V(\vec{k}_l, \thv_l) = \prod_{i=1}^n R_{k^i_l}(\theta_l^i)  \, ,
\end{equation}
where $R_{k_l^i}(\theta_l^i)$ is a rotation of the $i_{\rm th}$ qubit by an angle $\theta_l^i$ about the $k_l^i = x, y$ or $z$ axis. In the maximally expressive version of the ansatz the $x$, $y$ or $z$ rotation directions $\{ k_l^i \}$ for each qubit on each layer are chosen independently and with equal probability, and the rotation angles $\{ \theta_l^i \}$ are independently and randomly chosen in the range $0$ to $2 \pi$. Our numerics are implemented using TensorFlow Quantum~\cite{tfq}.

\medskip

\paragraph*{Circuit depth.} One of the simplest ways of reducing the expressiblity of an ansatz is reducing the depth $D$ of the circuit. It was shown in \cite{cerezo2020cost} that global costs with a hardware efficient ansatz experience barren plateaus irrespective of the depth of the circuit. However, 
local costs only exhibit barren plateaus for deep circuits ($D \in \Omega(\text{poly}(n)$) but are trainable for shallow circuits ($D \in \mathcal{O}(\text{log}(n)$). 

We obtain similar results here. As shown in Fig.~\ref{fig:VarGradPanels}(A), for the global cost the variance in the partial derivative is seemingly independent of the depth of the circuit and vanishes exponentially with the size of the system $n$. Conversely for local costs, as shown in Fig.~\ref{fig:VarGradPanels}(D), exponentially vanishing partial derivatives are observed for systems up to 12 qubits for depths $D \gtrapprox 100$. However shallow circuits $D \lessapprox 50$ exhibit an approximately constant scaling for $n \gtrapprox 8$. 

\medskip

\paragraph*{Correlating parameters.} A more sophisticated means of reducing the expressibility of the ansatz is to correlate the rotation angles~\cite{Volkoff2020BP}. Here we consider three different means of correlating parameters, as sketched in Fig.~\ref{fig:Ansatz}, and plot the corresponding variance in the cost partial derivative in the central panel of Fig.~\ref{fig:VarGradPanels}. In the first, shown in yellow, we correlate the qubits (but allow the angles to vary between layers), i.e., $k_l^i = k_l^{i'}$ and $\theta_l^i = \theta_l^{i'}$ for any two qubits $i$ and $i'$. In the second (plotted in green) we correlate the different layers (but not the qubits), i.e., $k^i_l = k^{i}_{l'}$ and $\theta_l^i = \theta_{l'}^i$ for any two layers $l$ and $l'$. Finally, as shown in blue, we correlate both the qubits and layers. In this case all the qubits rotate in same direction and by the same angle,  i.e., $k_l^i = k_{l'}^{i'}$ and $\theta_l^i = \theta_{l'}^{i'}$ for any two qubits $i$ and $i'$ and layers $l$ and $l'$. In other words, all parameters are correlated.  The data for only $y$ ($x$) rotations is indicated by the solid (dashed) lines respectively. 

In contrast to varying circuit depth, here we obtain similar results irrespective of whether a local or global cost is used. Correlating both the qubits and the layers results in the least expressive ansatz and correspondingly the largest variation in cost gradients is observed. Indeed, in this case the variance in the cost gradient is approximately constant. In contrast, correlating just the qubits, or just the layers, increases the cost gradients and reduces the scaling of the cost gradient with system size but an exponential scaling is still observed. 

\medskip

\paragraph*{Restricting rotation direction.} One might also consider reducing the expressibility of the ansatz by reducing the single qubit rotation gates to a subset of directions. We explore this in the right panel of Fig.~\ref{fig:VarGradPanels}. In blue we plot the variance when only rotations in a single direction, namely in the $x$ (dark blue) or $y$ (light blue) direction, are implemented. We do not plot the case when only $z$ rotations are implemented since in that case $U$ commutes with $H_L = \sigma_1^z \sigma_2^z$ and $H_G = \bigotimes_{i=1}^{n} \sigma_i^z$, and so the cost landscape is entirely flat.
For a local cost, reducing the expressibility of the ansatz by restricting to single direction rotations seemingly removes the exponential gradient scaling. However, for a global cost the scaling remains exponential.

\medskip

\paragraph*{Restricting rotation angles.} A final way to reduce the expressiblity of an ansatz is by reducing the range the rotation angles $\thv$ are chosen from. That is, choosing the $\theta_l^i$ in the range $[ \tilde{\theta}_l^i, \tilde{\theta}_l^i +2 \pi r ]$ where $ \tilde{\theta}_l^i $ is a fixed initialization point.
For $r = 1$ the ansatz explores the entire solution space but for $r < 1$ the ansatz is constrained to exploring a subset of the solution space where the rotation angles $\theta_l^i$ deviate from $\tilde{\theta}_l^i$ by at most $2 \pi r $.

\begin{figure}[t!]
    \centering
    \includegraphics[width=0.48\textwidth]{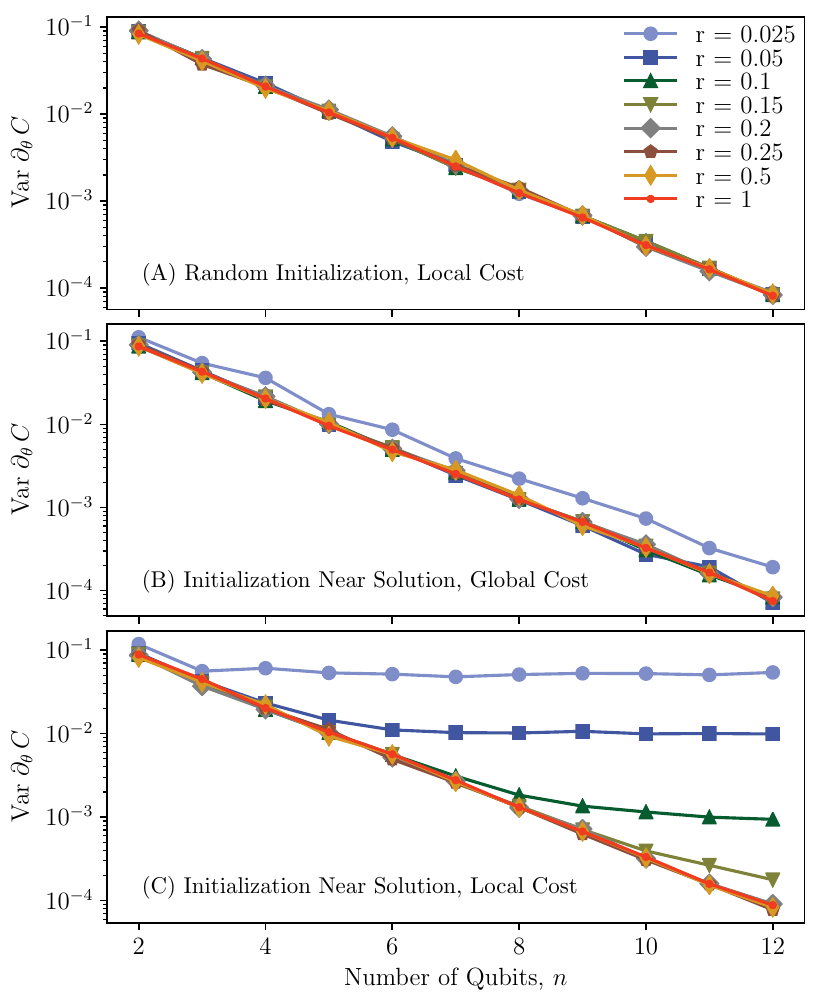}
    \caption{\textbf{Partial derivative scalings for restricted angle ranges.} The scaling of the variance in the partial derivative when the rotation angles $\theta_l^i$ are randomly chosen from the range $[ \tilde{\theta}_l^i, \tilde{\theta}_l^i +2 \pi r ]$, such that for $r = 1$ (red) the ansatz explores the entire solution space but for $r \ll 1$ (blue) the ansatz is constrained to exploring close to the initialization point defined by $\{\tilde{\theta}_l^i \}$. In (A), the angles $ \{ \tilde{\theta}_l^i \}$ are a fixed (randomly chosen) initialization point away from the solution (here we consider a local cost but the data for a global cost is essentially unchanged). In (B) and (C), which correspond to global and local costs respectively, the angles $ \tilde{\theta}_l^i = 0$ for all $l$ and $i$, which is close to the global minimum of the cost. 
    In all cases the derivative is taken with respect to $\theta_1^1$, the rotation angle of the first qubit in the first layer and the variance is taken over an ensemble of $1000$ unitaries. }\label{fig:VarGradAngleRange}
\end{figure}

However, with a little thought, it is clear that, in contrast to the previous three approaches we have discussed, restricting the rotation angles of the ansatz does not change the cost landscape but rather limits the region of the landscape explored by the ansatz. Thus, in general, reducing the rotation angles does not effect the cost gradients experienced.
This intuition is confirmed by the numerical results displayed in the top panel of Fig.~\ref{fig:VarGradAngleRange}. Here we randomly initialize the parameters by randomly choosing $\tilde{\theta}_l^i$ in the range $[0, 2 \pi]$. We find the the cost partial derivatives for different $r$ values perfectly overlap in this case, i.e., for a random initialization, restricting the ansatz to a limited range of rotation angles does not change the partial derivatives observed.

On the other hand, if the parameters are initialized close to the solution, varying $r$ has a substantial effect on the observed partial derivatives for local costs, and a reduced effect for global costs. This is seen in (B) and (C) of Fig.~\ref{fig:VarGradAngleRange} where we initialize to identity, i.e., pick $\tilde{\theta}_l^i = 0$ for all $i$, which is close to the solution for this simple problem. In this case, for $r$ close to 1 (as shown in red and yellow) the variance in the partial derivative again vanishes exponentially with $n$. However, for small angle ranges, $r \lessapprox 0.1$, as shown in blue, we find that the partial derivative of a local cost ceases to exhibit an exponential scaling. To some degree, a similar effect is displayed for global costs; however, the effect is reduced and is only visible in the data here for $r \approx 0.025$. 

This change in partial derivative scaling for small $r$ for initializations close to the solution is plausibly explained by the fact that the global minimum of costs exhibiting barren plateaus tend to sit within a steep and narrow gorge~\cite{cerezo2020cost}, as sketched in Fig.~\ref{fig:Schematic}(C). By initializing close to the solution we are likely to be initializing within the narrow gorge. In this case, when $r$ is close to 1 the ansatz still explores the entire cost landscape and therefore the variance in the partial derivative will be unchanged. However, for smaller $r$ the ansatz is constrained to the region around the the narrow gorge itself, and hence a larger variance in partial derivatives is observed.

\begin{figure}[t!]
    \centering
    \includegraphics[width=0.48\textwidth]{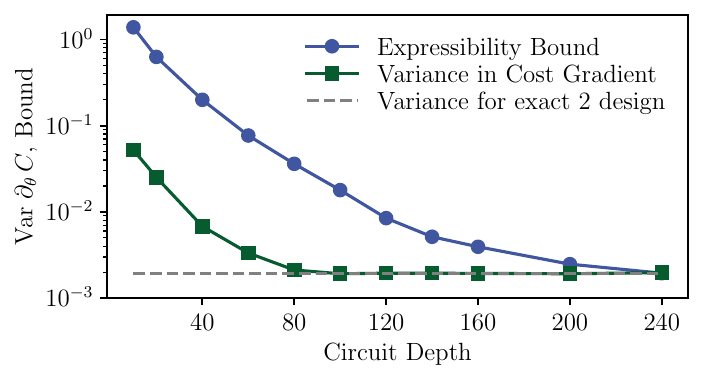}
    \caption{\textbf{Comparison of scaling of bound and gradients.} The scaling of the bound on the variance in the gradient (blue), Eq.~\eqref{eq:MainBoundLR}, and variance in the partial derivative (green) as a function of the ansatz depth for $n=8$ qubits. The dashed line indicates the predicted variance in the partial derivative for a perfect 2-design from Ref~\cite{BarrenPlateaus2018}. The derivative is taken with respect to $\theta_{D/2}^1$, the rotation angle of the first qubit in the middle layer ($D/2$) and the variance is taken over an ensemble of $1000$ unitaries. We chose to show the state and Hamiltonian dependent bound here, Eq.~\eqref{eq:MainBoundLR}, because as we are looking at the gradient with respect to $\theta_{D/2}^1$ this bound is tightest. }\label{fig:BoundTightness}
\end{figure}

\begin{figure*}[t!]
    \centering
    \includegraphics[width=\textwidth]{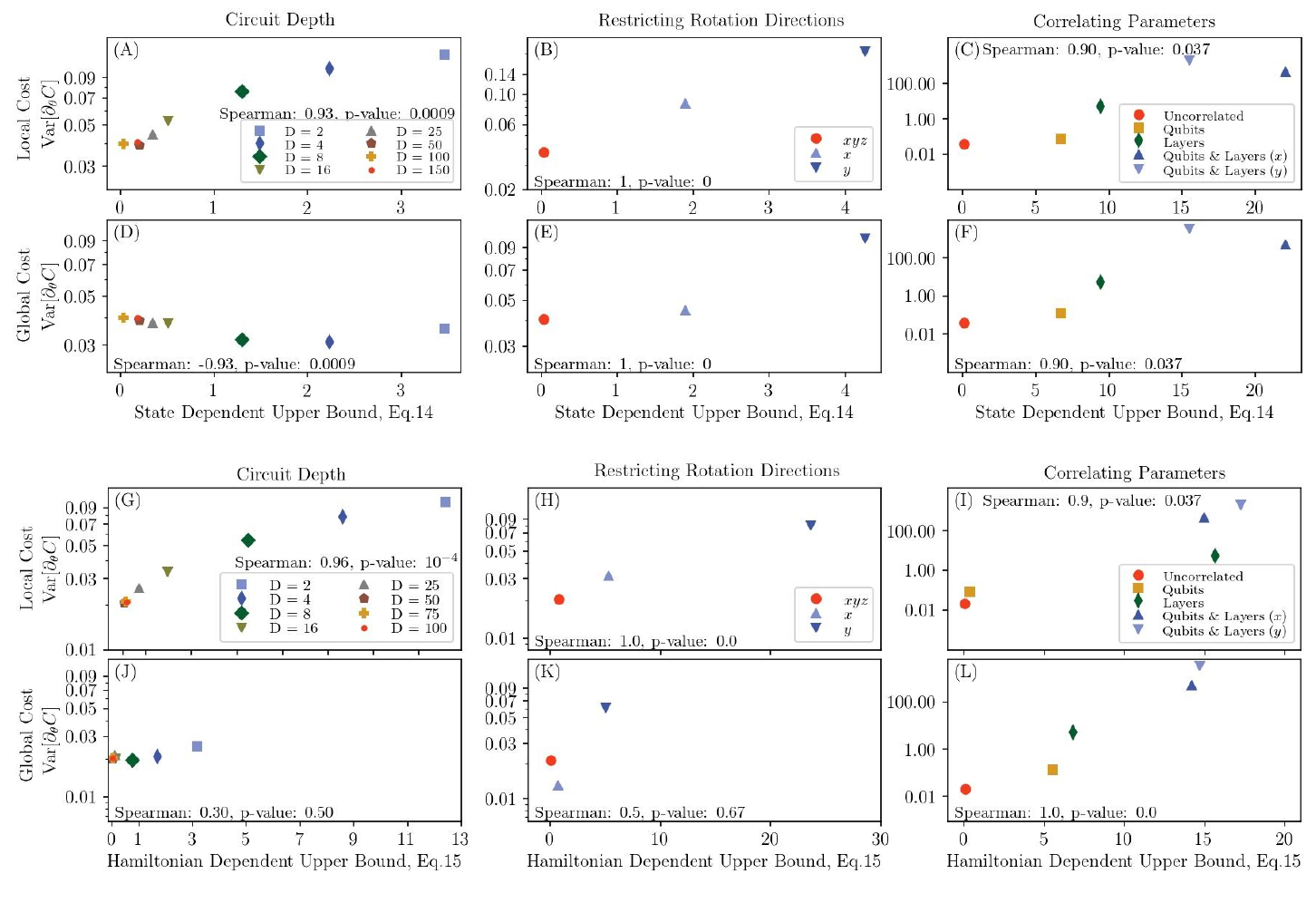}
    \caption{\textbf{Correlations between cost partial derivatives and bounds.} The variance in the partial derivative of a 2-local cost with $H = \sigma^z_1 \sigma^z_2$ (top) and global cost with $H = \prod_{i=1}^n \sigma^z_i$ (bottom) as a function of the state-dependent expressibility upper bound on the variance in the partial derivative specified by Eq.~\eqref{eq:MainBoundR} (A - F) and the Hamiltonian-dependent expressibility upper bound on the variance in the partial derivative specified by Eq.~\eqref{eq:MainBoundL} (G - L). In both cases $\rho = |\psi_0\rangle \langle \psi_0 |^{\otimes n} $ where $\ket{\psi_0} = \exp(-i(\pi/8)\sigma_Y) \ket{0}$. In the left panel we vary the circuit depth $D$ of a hardware efficient ansatz. In the right (middle) panel we consider the effect of correlating parameters (restricting the directions of rotation) of a hardware efficient ansatz with $D =100$ with the choices of correlations (rotations) indicated in the figure legend. In (A-F) the derivative is taken with respect to $\theta_1^1$ and in (G-L) the derivative is taken with respect to $\theta_D^1$. In all cases $n=4$ and the expressibility measures are estimated using an ensemble of $5000$ unitaries. }\label{fig:Bound1VersusGradsRho}
\end{figure*}

\medskip

\paragraph*{Outlook for ansatz design.}  Figure~\ref{fig:VarGradPanels} suggests that reducing the depth of a circuit and correlating parameters are the most effective strategies for amplifying the observed cost gradients.  However, the optimal solution, of course, may not lie within a shallow or highly correlated ansatz. When deep and/or uncorrelated circuits are required, as is expected to be the case for many problems of interest, then a perturbative strategy may instead be effective. That is, one could start the variational algorithm using a shallow, highly correlated ansatz and as the cost is iteratively minimized gradually grow the ansatz~\cite{VQSD,Grant2019initializationBP, cirstoiu2020variational} and decorrelate the parameters~\cite{Volkoff2020BP}. 

Restricting the angle range also appears to provide an effective strategy for increasing cost gradients, but for it to be practical it is necessary to initialize close to the solution. This, of course, requires either prior knowledge of an approximate solution to the problem at hand or an effective pre-training strategy to obtain such an approximate solution. The viability of either of these options warrants further investigation. 

\medskip

\paragraph*{Correlation and tightness of bounds.}

In Fig.~\ref{fig:Bound1VersusGradsRho} we study the correlation between the cost gradients and our upper bounds. To quantify this correlation we include the Spearman correlation coefficient~\cite{myers2013research}, as well as its corresponding p-value, which approximately gives the probability of uncorrelated data generating a Spearman coefficient at least as large as the one found. For local costs we obtain a Spearman value of at least 0.9 with a p-value of less than 0.05 in all cases, indicating a strong correlation between our upper bound and the actual variance in the gradient. The correlation is weaker in the case of global costs, highlighting that in the case of a global cost expressibility is not the only phenomenon that may induce a barren plateau.
This is to be expected given the results of Ref.~\cite{cerezo2020cost}, which show even very shallow and/or non-entangling circuits (i.e. highly inexpressive circuits) may exhibit barren plateaus when using global costs. For completeness, the results presented here are extended in Appendix~\ref{ap:ExpressVersusGrads}, where we study directly the correlation between cost gradient and the expressibility measures $\varepsilon_R^\rho$ and $\varepsilon_L^H$, with similar correlations observed.

Figure~\ref{fig:Bound1VersusGradsRho} additionally highlights that, as expected, the bounds are tightest for higher expressibility ans\"{a}tze but may be relatively loose for lower expressibilities. More specifically, in all cases considered here, the bounds are tight to within a couple of orders of magnitude, with the bounds tightest for ans\"{a}tze that are high depth, uncorrelated and use the full range of rotation directions\footnote{We note that for highly correlated circuits the variance in the gradient can in fact be larger than allowed by our bounds which were derived for uncorrelated circuits. In future work, it would be interesting to investigate whether our bounds can be generalised to account for such correlations.}. This phenomenon is more clearly demonstrated in Fig.~\ref{fig:BoundTightness} where we plot both the variance in the partial derivative of the cost and the Hamiltonian and state dependent expressiblity bound, Eq.~\eqref{eq:MainBoundLR}, as a function of ansatz depth for the 8 qubit local cost. The bound captures the qualitative behaviour of the cost gradients, decreasing with increased circuit depth. While moderately loose at low depths, the bound becomes tight for deep circuits.

\section{Discussion}

In this work, we extended the well-known barren plateau result. This result was restricted to ans\"{a}tze that form 2-designs~\cite{mcclean2018barren}, while we extended it to arbitrary ans\"{a}tze in our Theorems~\ref{thm:mainbounds} and \ref{corol:DiamondNormBounds}. In practice, this extension may prove to be quite useful, since many ans\"{a}tze of interest are not exact 2-designs but rather are some approximate notion of this~\cite{HarrowLowApproxDesigns09,  BrownFawzi2012,HarrowMehraban2018,  BrandaoHarrowHorodecki2016}. Our results can potentially provide useful bounds on the variance of the gradient in this realistic scenario of approximate 2-designs.

The key to our extension was to consider the expressiblity of the ansatz. This can be precisely defined in terms of the distance of the ensemble of unitaries accessible by the ansatz from being a 2-design. Hence, our extension linked two key properties of ans\"{a}tze: their expressiblity and their gradient magnitudes. Our bounds demonstrate that increasing the expressibility of an ansatz can result in smaller cost gradients. We believe that this connection is very interesting, and there is certainly much more to be explored along these lines. 
For example, it would be interesting to connect our findings to recent results on the role of the growth of entanglement in generating barren plateaus. In particular, since highly expressive ans\"{a}tze are necessarily highly entangling, our results would seem to imply those in Refs.~\cite{patti2020entanglement, marrero2020entanglement}.

To go beyond our bounds and look at the precise relation between expressiblity and gradients, we performed extensive numerics. We considered several different strategies by which one can vary the expressiblity. 
As highlighted in  Fig.~\ref{fig:Bound1VersusGradsRho} and Fig.~\ref{fig:BoundTightness}, we typically observed a strong correlation (especially for local cost functions) between the expressiblity and the variance of the gradient. However, the bounds are not perfectly tight. This may arise from the repeated use of the triangle and Cauchy-Schwarz inequalities in the derivation (Appendix~\ref{ap:Derivations}). Thus a natural question to ask is whether our bounds can be further tightened. Another direction would be to explore the nature of barren plateaus for global costs where the numerics suggest that the correlation between expressibility and cost gradients is weaker.

We remark that the numerical results presented here are necessarily problem specific, since they depend both on the choice in cost function and ansatz. Further work is required to ascertain the extent to which the trends observed here are universally observed. In particular it would be valuable to investigate whether any analytic results can be obtained to support them.

Nevertheless, there are several interesting trends shown in our numerics that even suggest potential strategies of avoiding or mitigating barren plateaus. As discussed above, correlating parameters and restricting rotation angles (especially when initializing near the solution) are two strategies that significantly mitigated barren plateaus in our numerics. Further exploring these and other strategies will be an important direction for future research.

\begin{acknowledgments}

ZH and PJC were supported by the Los Alamos National Laboratory (LANL) ASC Beyond Moore's Law project. KS was supported by the Laboratory Directed Research and Development (LDRD) program of LANL under project number 20190065DR. MC was initially supported by the LDRD program of LANL under project number 20180628ECR, and also supported by the Center for Nonlinear Studies at LANL. This work was also supported by the U.S. Department of Energy (DOE), Office of Science, Office of Advanced Scientific Computing Research, under the Accelerated Research in Quantum Computing (ARQC) program.

\end{acknowledgments}

\bibliography{refs} 

\onecolumngrid

\pagebreak

\appendix

\vspace{0.5in}

\begin{center}
	{\Large \bf Appendices} 
\end{center}

We begin by reviewing some definitions and prior results relevant for the rest of the appendices. We then provide proofs for the main results and theorems. 

\section{Preliminaries}

\noindent \textbf{Operator Norms.}
Let $\mathcal{D}(\mathcal{H})$ denote the set of density operators acting on a Hilbert space $\mathcal{H}$, i.e.,  those  that  are  positive  semi-definite  with  unit trace. Let $\mathcal{L}(\mathcal{H})$ denote the space of square linear operators acting on $\mathcal{H}$.
The trace norm or Schatten 1-norm $\Vert \Omega \Vert_1$ of an operator $\Omega\in \mathcal{L}(\mathcal{H})$ is defined as $\Vert \Omega \Vert_1 := \Tr[\vert \Omega \vert]$, where $\vert \Omega \vert := \sqrt{\Omega^{\dagger} \Omega}$. More generally, the Schatten $p$-norm of an operator $\Omega$ can be defined as $\Vert \Omega \Vert_p = (\Tr[\vert \Omega\vert^{p}])^{1/p}$, which satisfies $\Vert \Omega \Vert_p \leq \Vert \Omega\Vert_q$ for $p\geq q$. The diamond norm of a Hermiticity preserving linear map $\mathcal{S}_A$ is defined as 
\begin{align}\Vert \mathcal{S}_A \Vert_{\diamond} = \sup_{n} \sup_{\Omega_{AB}\neq 0} \frac{\Vert(\mathcal{S}_A\otimes \mathcal{I}_B^{(n)})(\Omega_{AB}) \Vert_1}{\Vert \Omega_{AB} \Vert_1},
\end{align}
where $\Omega_{AB} \in \mathcal{L}(\mathcal{H}_A \otimes \mathcal{H}_B)$ and $\mathcal{I}_B^{(n)}$ denote an identity channel acting on an $n$-dimensional system $B$. 
The diamond-norm distance $\Vert \mathcal{N} - \mathcal{M}\Vert_{\diamond}$ is a measure of the distinguishability of two quantum operations $\mathcal{N}$ and $\mathcal{M}$. 

\medskip

\noindent \textbf{Properties of the Haar measure.} Let $\mathcal{U}(d)$ denote the unitary group of degree $d=2^n$. Let $d\mu_H(V) = d\mu(V)$ be the volume element of the Haar measure, where $V\in \mathcal{U}(d)$. The volume of the Haar measure is finite: $\int_{\mathcal{U}(d)} d\mu(V) < \infty$. The Haar measure is uniquely defined up to a multiplicative constant factor. Let $d\zeta(V)$ be an invariant measure. Then there exists a constant $c$ such that $d\zeta(V) = c\cdot d\mu(V)$. The Haar measure is left- and right-invariant under the action of the unitary group of degree $d$, i.e., for any integrable function  $g(V)$, the following holds: \begin{align}\label{eq:LRinvar}
\int_{\mathcal{U}(d)} d\mu(V) g(W V) = \int_{\mathcal{U}(d)}d\mu(V) g(VW) = \int_{\mathcal{U}(d)} d\mu(V) g(V), 
\end{align}
where $W\in \mathcal{U}(d)$.

\medskip

\noindent \textbf{Symbolic integration.}
We recall formulas which allow for the symbolical integration with respect to the Haar measure on a unitary group~\cite{puchala2017symbolic}. For any $V\in \mathcal{U}(d)$ the following expressions are valid for the first two moments:  
\small
\begin{equation}\label{eq:delta}
\begin{aligned}
    \int d\mu(V)v_{\vec{i}\vec{j}}v_{\vec{p}\vec{k}}^*&=\frac{\delta_{\vec{i}\vec{p}}\delta_{\vec{j}\vec{k}}}{d}\,,   \\
\int d\mu(V)v_{\vec{i}_1\vec{j}_1}v_{\vec{i}_2\vec{j}_2}v_{\vec{i}_1'\vec{j}_1'}^{*}v_{\vec{i}_2'\vec{j}_2'}^{*}&=\frac{\delta_{\vec{i}_1\vec{i}_1'}\delta_{\vec{i}_2\vec{i}_2'}\delta_{\vec{j}_1\vec{j}_1'}\delta_{\vec{j}_2\vec{j}_2'}+\delta_{\vec{i}_1\vec{i}_2'}\delta_{\vec{i}_2\vec{i}_1'}\delta_{\vec{j}_1\vec{j}_2'}\delta_{\vec{j}_2\vec{j}_1'}}{d^2-1}
-\frac{\delta_{\vec{i}_1\vec{i}_1'}\delta_{\vec{i}_2\vec{i}_2'}\delta_{\vec{j}_1\vec{j}_2'}\delta_{\vec{j}_2\vec{j}_1'}+\delta_{\vec{i}_1\vec{i}_2'}\delta_{\vec{i}_2\vec{i}_1'}\delta_{\vec{j}_1\vec{j}_1'}\delta_{\vec{j}_2\vec{j}_2'}}{d(d^2-1)}\,,
\end{aligned}
\end{equation}
\normalsize
where $v_{\vec{i}\vec{j}}$ are the matrix elements of $V$. Assuming $d=2^n$, we use the notation $\vec{i} = (i_1, \dots i_n)$ to denote a bitstring of length $n$ such that $i_1,i_2,\dotsc,i_{n}\in\{0,1\}$. 

\medskip

\noindent \textbf{Useful Identities.} 
We use the following identities, which can be derived using Eq.~\eqref{eq:delta} (see~\cite{cerezo2020cost} for a review):
\small
\begin{align}
    \int d \mu(W) \operatorname{Tr}\left[W A W^{\dagger} B\right]&=\frac{\operatorname{Tr}[A] \operatorname{Tr}[B]}{d},\label{eq:identity1}\\
    \int d \mu(W) \operatorname{Tr}\left[W A W^{\dagger} B W C W^{\dagger} D\right] &=\frac{\operatorname{Tr}[A] \operatorname{Tr}[C] \operatorname{Tr}[B D]+\operatorname{Tr}[A C] \operatorname{Tr}[B] \operatorname{Tr}[D]}{d^{2}-1}  -\frac{\operatorname{Tr}[A C] \operatorname{Tr}[B D]+\operatorname{Tr}[A] \operatorname{Tr}[B] \operatorname{Tr}[C] \operatorname{Tr}[D]}{d\left(d^{2}-1\right)},\label{eq:identity2}\\
        \int d \mu(W) \operatorname{Tr}\left[W A W^{\dagger} B\right] \operatorname{Tr}\left[W C W^{\dagger} D\right]& = \frac{\operatorname{Tr}[A] \operatorname{Tr}[B] \operatorname{Tr}[C] \operatorname{Tr}[D]+\operatorname{Tr}[A C] +\operatorname{Tr}[B D]}{d^{2}-1}-\frac{\operatorname{\Tr}[A C] \operatorname{Tr}[B] \operatorname{Tr}[D]+\operatorname{Tr}[A] \operatorname{Tr}[C] \operatorname{Tr}[B D]}{d\left(d^{2}-1\right)},\label{eq:identity3}
\end{align}
\normalsize
    where $A, B, C$, and $D$ are linear operators on a $d$-dimensional Hilbert space. 

 Let $A \in \mathcal{L}(\mathcal{H})$ and $B\in \mathcal{L}(\mathcal{H}')$. Then the following identity holds: 
\begin{align}\label{eq:trace-tprod}
    \Tr[A]\Tr[B] = \Tr[A \otimes B].
\end{align}

Let $A, B \in \mathcal{L}(\mathcal{H})$, where $\mathcal{H}$ is a $d^2$-dimensional Hilbert space. 
Then from Eq.~\eqref{eq:delta}, we derive the following integral:
\begin{equation}\label{eq:Int}
\begin{aligned}
        \int d\mu(U) \Tr[A U^{\otimes2}  B U^{\dagger \otimes2} ] = \frac{\Tr[A] \Tr[B] + \Tr[A W] \Tr[B W] }{d^2-1}- \frac{\Tr[A W] \Tr[B] + \Tr[A] \Tr[B W]}{d(d^2-1)}
\end{aligned}
\end{equation}
where $W$ is the subsystem swap operator, i.e., $W \ket{i}\ket{j} = \ket{j}\ket{i}$. 

\section{Definitions of Expressibility}

In broad terms a parameterized quantum circuit can be considered expressive if the circuit can be used to uniformly explore the unitary group $\mathcal{U}(d)$. Thus, the expressiblity of a circuit can be defined in terms of 
the following super-operator 
\begin{align}
    \mathcal{A}^{(t)}_{\mathbb{U}}(\cdot):= \int_{\mathcal{U}(d)} d\mu(V)  V^{\otimes t}  ( \,\cdot \, ) (V^\dagger)^{\otimes t}  -  \int_{\mathbb{U}} d U \, U^{\otimes t} ( \,\cdot\, ) (U^\dagger)^{\otimes t} \, 
\end{align}
where $d\mu(V)$ is the volume element of the Haar measure and $dU$ is the volume element corresponding to the uniform distribution over $\mathbb{U}$. If $ \mathcal{A}^{(t)}_{\mathbb{U}}(X) =  0$ for all operators $X$ then the averaging over elements of $\mathbb{U}$ agrees with averaging over the Haar distribution up to the $t$-th moment. In this case $\mathbb{U}$ is said to form a \textit{$t$-design}. For our purposes it suffices to consider the behavior of $\mathcal{A}^{(t)}_{\mathbb{U}}$ for $t=2$. Henceforth, we drop the t-superscript and denote $\mathcal{A}^{(2)}_{\mathbb{U}}(\cdot)$ as $\mathcal{A}_{\mathbb{U}}(\cdot)$. In the context of minimizing a generic cost $C$ of the form specified by Eq.~\eqref{eq:GenCost}, we are interested in the quantities
\begin{align}
    &\varepsilon_{\mathbb{U}}^\rho := || \AC_{{\mathbb{U}}}( \rho^{\otimes 2}) ||_2 \\
    &\varepsilon_{\mathbb{U}}^H := || \AC_{{\mathbb{U}}}( H^{\otimes 2}) ||_2 \, .
\end{align}

The quantities $\varepsilon_{\mathbb{U}}^\rho $ and $\varepsilon_{\mathbb{U}}^H $ may be more readily computed by relating them to a generalization of the frame potential. To demonstrate how, let us first recall that the frame potential~\cite{ChaosByDesign,ExpressibilityNakaji2020} of an ensemble $\mathbb{U}$ may be defined as
\begin{equation}
      \mathcal{F}_{\mathbb{U}} := \int_{\mathbb{U}} \int_{\mathbb{U}} dU dV |\bra{0} (U V^\dagger) \ket{0} |^4 \, ,
\end{equation}
where $dU$ and $dV$ are volume elements corresponding to the distribution over $\mathbb{U}$. 
We then note that the quantity $|| \mathcal{A}_{{\mathbb{U}}}( \vert 0 \rangle \langle 0 \vert) ||_2^2$ can be rewritten in terms of $\mathcal{F}_{{\mathbb{U}}}$ as follows 
\begin{equation}
\begin{aligned}\label{eq:AtoFP}
          || \mathcal{A}_{{\mathbb{U}}}( \vert 0 \rangle \langle 0 \vert) ||_2^2  &= \bigg| \bigg| \int_{\mathcal{U}(d)} d\mu(V) \, V^{\otimes 2}  ( \vert 0 \rangle \langle 0 \vert ) (V^\dagger)^{\otimes 2}  -  \int_{\mathbb{U}} d U \, U^{\otimes 2} ( \vert 0 \rangle \langle 0 \vert ) (U^\dagger)^{\otimes 2} \bigg| \bigg|_2^2 \\ 
          &= \int_{\mathbb{U}} \int_{\mathbb{U}} dU dV |\bra{0} (U V^\dagger) \ket{0} |^4  -  2 \int_{\mathcal{U}(d)} \int_{\mathbb{U}} d\mu(V) dU |\bra{0} (U V^\dagger) \ket{0} |^4  +  \int_{\mathcal{U}(d)} \int_{\mathcal{U}(d)} d\mu(U) d\mu(V) |\bra{0} (U V^\dagger) \ket{0} |^4 \\
          &= \int_{\mathbb{U}} \int_{\mathbb{U}} dU dV |\bra{0} (U V^\dagger) \ket{0} |^4  -  \int_{\mathcal{U}(d)} \int_{\mathcal{U}(d)} d\mu(U) d\mu(V) |\bra{0} (U V^\dagger) \ket{0} |^4 \\
          &= \mathcal{F}_{\mathbb{U}} - \mathcal{F}_{\rm Haar} \, ,
\end{aligned}
\end{equation}
where we use the left and right invariance of the Haar measure, i.e. Eq.~\eqref{eq:LRinvar}, as in Ref.~\cite{ChaosByDesign}, and where we defined
\begin{equation}
    \mathcal{F}_{\rm Haar} := \int_{\mathcal{U}(d)} \int_{\mathcal{U}(d)} d\mu(U) d\mu(V) |\bra{0} (U V^\dagger) \ket{0} |^4  = \frac{1}{(2^n+1)2^{n-1}}  \, .
\end{equation}

In the context of the expressibility of a VQA we are interested in the more general quantity $|| \mathcal{A}_{{\mathbb{U}}}( X^{\otimes 2} ) ||_2 $ where $X$ is a quantum state $\rho$ or Hamiltonian $H$. Following the same approach as in Eq.~\eqref{eq:AtoFP}, we note that $|| \mathcal{A}_{{\mathbb{U}}}( X^{\otimes 2} ) ||_2 $ can be rewritten as
\begin{equation}
\begin{aligned}
          || \mathcal{A}_{{\mathbb{U}}}( X^{\otimes 2} ) ||_2 
          &= \sqrt{ \mathcal{F}^{(X)}_{\mathbb{U}} - \mathcal{F}^{(X)}_{\rm Haar} } \, ,
\end{aligned}
\end{equation}
where we have defined the operator dependent frame-potential as
\begin{equation}
    \mathcal{F}^{(X)}_{\mathbb{U}} :=  \int_{\mathbb{U}} \int_{\mathbb{U}} dU dV \Tr [ X U^\dagger V X V^\dagger U  ]^2 \, 
\end{equation}
and 
\begin{equation}
    \mathcal{F}^{(X)}_{\rm Haar} :=  \int_{\mathcal{U}(d)} \int_{\mathcal{U}(d)} d\mu(U) d\mu(V) \Tr [ X U^\dagger V X V^\dagger U  ]^2 \, .
\end{equation}
The latter can be evaluated using Eq.~\eqref{eq:identity3} to give \begin{equation}\label{eq:FPsHaar}
    \mathcal{F}^{(X)}_{\rm Haar} =  \frac{ \Tr[X]^4 + \Tr[X^2]^2  }{2^{2n} -1 } - \frac{2 \Tr[X^2] \Tr[X]^2 }{2^{n}(2^{2n} - 1)} \, . 
\end{equation}
Thus our expressibility measures can be related to state and Hamiltonian dependent frame potentials via 
\begin{align}\label{eq:FPs}
    &\varepsilon_{\mathbb{U}}^\rho := || \AC_{{\mathbb{U}}}( \rho^{\otimes 2}) ||_2 = \sqrt{ \mathcal{F}^{(\rho)}_{\mathbb{U}} - \mathcal{F}^{(\rho)}_{\rm Haar}} \\
    &\varepsilon_{\mathbb{U}}^H := || \AC_{{\mathbb{U}}}( H^{\otimes 2}) ||_2  = \sqrt{ \mathcal{F}^{(H)}_{\mathbb{U}} - \mathcal{F}^{(H)}_{\rm Haar}}\, .
\end{align}
We will use these expressions to evaluate the expressiblity of different ans\"{a}tze in Appendix~\ref{ap:ExpressVersusGrads}

\section{Proof for Eq.~\eqref{eq:GradUnbiased}}\label{ap:Unbiased}

For a random layered parametrized ansatz of the form Eq.~\eqref{eq:LayeredAnsatz} and Eqs.~\eqref{eq:structure}--\eqref{eq:ulur}, and the generic cost defined in Eq.~\eqref{eq:GenCost}, we now show that $\langle \partial_k C \rangle_{\mathbb{U}} = 0 $ for all $k$ and therefore the cost landscape is unbiased. 

To do so, let us first note that the cost function can be expressed as 
\begin{align}
    C = \Tr[ U_k(\theta_k) \widetilde{\rho} U_k(\theta_k)^{\dagger} \widetilde{H}].
\end{align}
where we introduce the shorthand 
\begin{align}
\widetilde{\rho} &= W_k \left(\prod_{j=1}^{k-1} U_j(\theta_j) W_j\right)\rho\left(\prod_{j=1}^{k-1} U_j(\theta_j) W_j\right)^{\dagger} W_k^{\dagger},\\ 
\widetilde{H} &= U_L(\vec{\theta})^{\dagger}HU_L(\vec{\theta}). 
\end{align}
This rewriting emphasises the dependence of $C$ on $U_k(\theta_k)$, the rotation we are taking the partial derivative with respect to, by associating $U_L$ with the Hamiltonian $H$ and $\left(\prod_{j=1}^{k-1} U_j(\theta_j) W_j\right)$ with $\rho$.

It follows that that 
\begin{align}
    \partial_{k}C &= -i \Tr[V_k U_k(\theta_k) \widetilde{\rho} U_k(\theta_k)^{\dagger} \widetilde{H}] + i\Tr[U_k(\theta_k) \widetilde{\rho} U_k(\theta_k)^{\dagger} V_k\widetilde{H}] \\
    & = -i \Tr[V_k (\cos(\theta_k) - i \sin(\theta_k)V_k) \widetilde{\rho} (\cos(\theta_k)  + i \sin(\theta_k)V_k) \widetilde{H}] \nonumber \\
    & \qquad \qquad+ i\Tr[(\cos(\theta_k) - i \sin(\theta_k)V_k) \widetilde{\rho} (\cos(\theta_k) + i \sin(\theta_k)V_k) V_k\widetilde{H}]\\
    & = -i \left((\cos(\theta_k)^2 - \sin(\theta_k)^2)(\Tr[V_k \widetilde{\rho} \widetilde{H}] - \Tr[\widetilde{\rho} V_k \widetilde{H}])+i\sin(2\theta_k)(\Tr[V_k \widetilde{\rho} V_k \widetilde{H}] - \Tr[\widetilde{\rho}\widetilde{H}]) \right)~.
\end{align}
Since $\int_0^{2\pi} \sin(2\theta_k)=0$ and $\int_0^{2\pi} \cos(\theta_k)^2 =\int_0^{2\pi} \sin(\theta_k)^2$, uniform averaging of $\partial_k C$ over $\theta_k$ leads to 
\begin{align}
  \frac{1}{2\pi}\int_0^{2\pi} d\theta_k \partial_k C &= 0,
\end{align}
which implies that $\langle \partial_k C \rangle_\mathbb{U}=0$.

\section{Variance of the partial derivative derivation}\label{ap:Derivations}

For a random layered parametrized ansatz of the form Eqs.~\eqref{eq:LayeredAnsatz} and \eqref{eq:structure}--\eqref{eq:ulur}, and the generic cost defined in Eq.~\eqref{eq:GenCost}, then since
\begin{equation}\label{eq:DerivationCost}
    \partial_k U(\vec{\theta}) = - i U_L V_k U_R
\end{equation}
where $U_L$ and $U_R$ are defined in Eq.~\eqref{eq:ulur}, 
it follows that the partial derivative of the cost can be written as 
\begin{equation}\label{eq:gradCost}
    \partial_k C := \frac{\partial C}{\partial \theta_k } = i  \Tr[ U_R \rho U_R^\dagger [V_k, U_L^\dagger H U_L] ] \, .
\end{equation}

Since the average derivative of the cost vanishes, as discussed in Appendix~\ref{ap:Unbiased}, its variance is given by
\begin{equation}\label{eq:Var}
    \text{Var} \, \partial_k C = \langle (\partial_k C)^2 \rangle_{\mathbb{U}} \, .
\end{equation}
Eq.~\eqref{eq:gradCost} and Eq.~\eqref{eq:Var} provide the starting point to derive the bounds Eqs.~\eqref{eq:MainBoundR}--\eqref{eq:MainBoundLR} and Eqs.~\eqref{eq:boundRd}--\eqref{eq:MainBoundLRd}.

\subsection{Bound in Eq.~\eqref{eq:MainBoundR}.}

Note that two different ensembles $\mathbb{U}_L$ and $\mathbb{U}_R$ can be generated using $U_L(\vec{\theta})$ and $U_R(\vec{\theta})$, respectively, as defined in  Eq.~\eqref{eq:structure}. Let $dU_L$ and $dU_R$ denote volume elements corresponding distributions over $\mathbb{U}_L$ and $\mathbb{U}_R$, respectively. Since $\mathbb{U}_L$ and $\mathbb{U}_R$ are independent, from the definition of $dU$ and from Eq.~\eqref{eq:structure}, we get that $dU = dU_L dU_R$. 

Then by substituting Eq.~\eqref{eq:gradCost} into Eq.~\eqref{eq:Var} and using Eq.~\eqref{eq:trace-tprod}, we get
\begin{equation}\label{eq:variance-first-step}
    \Var \partial_k C = - \int_{\mathbb{U}_L} d U_L \int_{\mathbb{U}_R} d U_R \Tr[U_R^{\otimes 2} \rho^{\otimes 2} {U_R^\dagger}^{\otimes 2} X_{Lk}^{\otimes 2}]
\end{equation}
where 
\begin{align}\label{eq:xlk}
X_{Lk} := [V_k, U_L^\dagger H U_L].
\end{align} 

Next we substitute in $\mathcal{A}_R(\rho^{\otimes 2})$ to give 
\begin{equation}
\begin{aligned}
    \Var \partial_k C &= - \int_{\mathbb{U}_L} d U_L \int_{\mathcal{U}(d)} d\mu(U) \Tr[U_{\rm Haar}^{\otimes 2} \rho^{\otimes 2} U_{\rm Haar}^{\dagger \otimes 2} X_{Lk}^{\otimes 2}] + \int_{\mathbb{U}_L} d U_L  \Tr[\mathcal{A}_R(\rho^{\otimes 2}) X_{Lk}^{\otimes 2}] \\ 
    &= \Var_R \partial_k C +  \int_{\mathbb{U}_L} d U_L \Tr[\mathcal{A}_R(\rho^{\otimes 2}) X_{Lk}^{\otimes 2}] \, 
\end{aligned}
\end{equation}
where in the second line we use the explicit definition of $\Var_R \partial_k C$, the variance in the partial derivative of the cost when $\mathbb{U_R}$ forms a 2-design, i.e.,
\begin{equation}\label{eq:variance-first-step}
    \Var_R \partial_k C := - \int_{\mathbb{U}_L} d U_L \int_{\mathcal{U}(d)} d\mu(U) \Tr[U_{\rm Haar}^{\otimes 2} \rho^{\otimes 2} U_{\rm Haar}^{\dagger \otimes 2} X_{Lk}^{\otimes 2}] \, .
\end{equation}
Rearranging we are left with 
\begin{equation}
    | \text{Var} \, \partial_k C - \text{Var}_R \partial_k C | \leq  \bigg| \int_{\mathbb{U}_L} d U_L \Tr[\mathcal{A}_R(\rho^{\otimes 2}) X_{Lk}^{\otimes 2}] \bigg| \, 
\end{equation}
which on using the triangle inequality followed by the Cauchy-Schwarz inequality reduces to 
\begin{equation}
\begin{aligned}\label{eq:DiffVarR}
        | \text{Var} \, \partial_k C - \text{Var}_R \partial_k C | &\leq \int_{\mathbb{U}_L} d U_L |  \Tr[\mathcal{A}_R(\rho^{\otimes 2}) X_{Lk}^{\otimes 2}] | \\
        &\leq  \int_{\mathbb{U}_L} d U_L || X_{Lk}^{\otimes 2}||_2 || \mathcal{A}_R(\rho^{\otimes 2}) ||_2\, .
\end{aligned}
\end{equation}
The term $|| X_{Lk}^{\otimes 2}||_2$ can be bounded as follows. First we note that $X_{Lk}^\dagger = - X_{Lk}$, which implies that
\begin{equation}\label{eq:xlk-simplification}
    || X_{Lk}^{\otimes 2}||_2 = \sqrt{\Tr[X_{Lk}^{\otimes 2}X_{Lk}^{\otimes 2}]} = \sqrt{\Tr[X_{Lk}^{2} \otimes X_{Lk}^{2}]} = \vert \Tr[X_{Lk}^2]\vert = \vert \Tr[[V_k, U_L^\dagger H U_L]^2] \vert\, .
\end{equation}
Let $A = V_k$ and $B= U_L^{\dagger} H U_L$. Since $A$ and $B$ are Hermitian, from the triangle inequality and the Cauchy-Schwarz inequality, we get 
\begin{equation}
\vert \Tr[ [A, B]^2 ]\vert = 2 \vert\Tr[ABAB] -  \Tr[A^2 B^2] \vert \leq 2 [\vert\Tr[ABAB]\vert+\vert\Tr[B^2]\vert] \leq 2 \sqrt{\Tr[ABAABA] \Tr[B^2]}  + 2 \vert \Tr[B^2]\vert = 4 \vert \Tr[B^2] \vert  \, .
\end{equation}
Therefore, we find that 
\begin{equation}
    || X_{Lk}^{\otimes 2}||_2 \leq 4 \Tr[(U_L^\dagger H U_L)^2] = 4\Tr[H^2] = 4 || H||_2^2 \, .
\end{equation}

Hence the bound takes the form
\begin{equation}\label{eq:first-bound-hcost}
    | \text{Var} \, \partial_k C - \text{Var}_R \partial_k C | \leq 4  \int_{\mathbb{U}_L} d U_L   || \mathcal{A}_R(\rho^{\otimes 2}) ||_{2} ||H||_{2}^2  =  4 || \mathcal{A}_R(\rho^{\otimes 2}) ||_{2} ||H||_{2}^2 \, ,
\end{equation}
which completes the proof. 

\medskip

\paragraph*{Extension to generalized cost.} This result can be further extended to cost functions of the following form 
\begin{align}\label{eq:gen-cost}
    C(\vec{\theta}) = \sum_m \Tr[H_m U(\vec{\theta}) \rho_m U(\vec{\theta})^{\dagger}]~,
\end{align}
for which the derivative with respect to the parameter $\theta_k$ can be written as 
\begin{align}
    \partial_k C = i \sum_m\Tr[U_R \rho_m U_R^{\dagger}[V_k, U_L^{\dagger} H_m U_L]]~.
\end{align}

Therefore, from Eq.~\eqref{eq:variance-first-step} it follows that 
\begin{align}
    \text{Var}\partial_k C = - \sum_{m,n} \int_{\mathbb{U}_L} dU_L \int_{\mathbb{U}_R} dU_R \Tr[U_R^{\otimes 2} (\rho_m\otimes \rho_n) (U_R^{\dagger})^{\otimes 2} X^m_{Lk}\otimes X^n_{Lk}],
\end{align}
where $X_{Lk}^m$ is defined in Eq.~\eqref{eq:xlk} with $H= H_m$.  

After substituting $\mathcal{A}_R(\rho_j \otimes \rho_k)$, we get 
\begin{align}
    \text{Var}\partial_k C = \text{Var}_R \partial_k C + \sum_{m,n} \int_{\mathbb{U}_L} dU_L \Tr[\mathcal{A}_{R}(\rho_m \otimes \rho_n)(X_{Lk}^m \otimes X_{Lk}^n)],
\end{align}
which implies that 
\begin{align}
| \text{Var} \, \partial_k C - \text{Var}_R \partial_k C | &\leq \sum_{m,n}\int_{\mathbb{U}_L} d U_L \vert \Tr[\mathcal{A}_{R}(\rho_m \otimes \rho_n)(X_{Lk}^m \otimes X_{Lk}^n)]\vert \\
&  \leq \sum_{m,n}\int_{\mathbb{U}_L} dU_L \Vert \mathcal{A}_{R}(\rho_m \otimes \rho_n) \Vert_2 \Vert (X_{Lk}^m \otimes X_{Lk}^n) \Vert_2\\
& \leq \sum_{m,n} \Vert\mathcal{A}_{R}(\rho_m \otimes \rho_n) \Vert_2 \sqrt{\Tr[(X_{Lk}^{m})^2]\Tr[(X_{Lk}^n)^2]}\\
& \leq 4\sum_{m,n}\Vert\mathcal{A}_{R}(\rho_m \otimes \rho_n) \Vert_2 \Vert H_m\Vert_2 \Vert H_n \Vert_2~,\label{eq:ar-rhojk-hjk}
\end{align}
where we used steps similar to those used in deriving Eqs.~\eqref{eq:xlk-simplification}--\eqref{eq:first-bound-hcost}.

\subsection{Bound in Eq.~\eqref{eq:MainBoundL}.}
Substituting Eq.~\eqref{eq:gradCost} into Eq.~\eqref{eq:Var}, using Eq.~\eqref{eq:trace-tprod}, and the cyclicity of the trace operation, we find that
\begin{equation}
    \Var \partial_k C = \int_{\mathbb{U}_L} d U_L \int_{\mathbb{U}_R} d U_R \Tr[U_L^{\dagger \otimes 2} H^{\otimes 2} {U_L}^{\otimes 2} Y_{Rk}^{\otimes 2}]
\end{equation}
where $Y_{Rk} := [U_R \rho U_R^\dagger, V_k]$. The rest of the derivation proceeds in the same manner as for the bound in Eq.~\eqref{eq:MainBoundR}. 

\medskip

\paragraph*{Extension to generalized cost.} 
Similar to Eq.~\eqref{eq:ar-rhojk-hjk}, the bound in Eq.~\eqref{eq:MainBoundL} can be extended for the cost functions of the form in Eq.~\eqref{eq:gen-cost}. In particular, we find that 
\begin{align}\label{eq:al-rhojk-hjk}
    \vert \text{Var}\partial_kC - \text{\Var}_L \partial_k C\vert \leq 4 \sum_{m,n} \Vert \mathcal{A}_{L}(H_m \otimes H_n) \Vert_2 \Vert \rho_m\Vert_2 \Vert \rho_n \Vert_2~.
\end{align}

\subsection{Bound in Eq.~\eqref{eq:MainBoundLR}.}

To derive Eq.~\eqref{eq:MainBoundLR} we start by substituting Eq.~\eqref{eq:gradCost} into Eq.~\eqref{eq:Var}, using Eq.~\eqref{eq:trace-tprod}, and the cyclicity of the trace operation to find that
\begin{equation}
    \Var \partial_k C = - \int_{\mathbb{U}_L} d U_L \int_{\mathbb{U}_R} d U_R \Tr[\rho_R^{\otimes 2} ( V_k^{\otimes 2} H_L^{\otimes 2} + H_L^{\otimes 2} V_k^{\otimes 2} - 2 (V_k \otimes \I) H_L^{\otimes 2} (\I \otimes V_k) )]
\end{equation}
where we have introduced the short hand $\rho_R := U_R \rho U_R^\dagger$ and $H_L := U_L^\dagger H U_L$. Next we substitute in $\mathcal{A}_L(H^{\otimes 2})$ and $\mathcal{A}_R(\rho^{\otimes 2})$ to find that the variance is given by
\begin{equation}\label{eq:VarLRinter}
    \Var \partial_k C = \Var_{L,R} \partial_k C - \Tr[\mathcal{A}_R(\rho^{\otimes 2}) Z_{Lk}] + I_1 + I_2 \, .
\end{equation}
Here we defined
\begin{align}
Z_{xk} := ( V_k^{\otimes 2} \mathcal{A}_x(\omega_x) + \mathcal{A}_x(\omega_x) V_k^{\otimes 2} - 2 (V_k \otimes \I) \mathcal{A}_x(\omega_x) (\I \otimes V_k) )
\end{align}
for $x = L$ and $x = R$, and where $\omega_R = \rho$ and $\omega_L = H$. The integrals $I_1$ and $I_2$ are given by
\begin{equation}
    \begin{aligned}
    &I_1 = \int_{\mathcal{U}(d)} d\mu(U) \Tr[ Z_{Lk} \tilde{\rho}^{\otimes 2} ] \\ 
    &I_2 =  \int_{\mathcal{U}(d)} d\mu(U) \Tr[ \mathcal{A}_R(\rho^{\otimes 2})  ( V_k^{\otimes 2} \tilde{H}^{\otimes 2} + \tilde{H}^{\otimes 2} V_k^{\otimes 2} - 2 (V_k \otimes \I) \tilde{H}^{\otimes 2} (\I \otimes V_k) ] \, .
    \end{aligned}
\end{equation}
with $\tilde{\rho} = U \rho U^\dagger$ and  $\tilde{H} = U^\dagger H U$. 

The integrals $I_1$ and $I_2$ can be evaluated using Eq.~\eqref{eq:Int} as follows:
\begin{equation}\label{eq:I1}
\begin{aligned}
     I_1  &= \frac{1}{d^2 - 1} \Tr[Z_{Lk} W] \Tr[\rho^2] - \frac{1}{d(d^2-1)}\Tr[Z_{Lk} W]~,\\ 
    I_2 &= \frac{1}{d^2 - 1} \Tr[Z_{Rk} W] \Tr[H^2] - \frac{1}{d(d^2-1)}\Tr[Z_{Rk} W] \Tr[H]^2,
\end{aligned}
\end{equation}
where we used the fact that $\Tr[Z_{Lk}] = \Tr[Z_{Rk}] = 0$, $\Tr[\rho^{\otimes 2} W] = \Tr[\rho^2]$, and $\Tr[H^{\otimes 2} W] = \Tr[H^2]$. 

Substituting these integrals, Eq.~\eqref{eq:I1}, back into Eq.~\eqref{eq:VarLRinter} and then using the triangle inequality gives
\begin{equation}\label{eq:VarDiffLR}
    | \text{Var} \, \partial_k C - \text{Var}_{R, L} \partial_k C | \leq  \frac{1}{d^2-1}((\Tr[\rho^2]-1/d)|\Tr[Z_{Lk} W]| + (\Tr[H^2]-\Tr[H]^2/d) |\Tr[Z_{Rk} W]|) + |\Tr[\mathcal{A}_R(\rho^{\otimes 2}) Z_{L, k}]| \, .
\end{equation}
Using Cauchy-Schwarz this reduces to 
\begin{equation}
    | \text{Var} \, \partial_k C - \text{Var}_{R, L} \partial_k C | \leq  \frac{d}{d^2 - 1 }((||\rho||_2^2-1/d)|| Z_{Lk}||_2  + (||H||_2^2-\Tr[H]^2/d) || Z_{Rk}||_2 )+ || \mathcal{A}_R(\rho^{\otimes 2}) ||_2 ||Z_{Lk}||_2 \, .
\end{equation}
where we have used $|| W ||_2 = d$. Finally, by expanding $|| Z_{xk} ||_2$, using the triangle inequality and the fact that $V_k^2 = \I$ we find that
\begin{equation}
    || Z_{xk} ||_2 \leq 4 || \mathcal{A}_x(\omega_x) ||_2 \, 
\end{equation}
for $x = L$ and $x = R$, and where $\omega_R = \rho$ and $\omega_L = H$.
Thus we are left with 
\begin{align}
    | \text{Var} \,&\partial_k C - \text{Var}_{R, L} \partial_k C |\nonumber \\ 
    &\leq 4 || \mathcal{A}_R(\rho^{\otimes 2}) ||_{2} || \mathcal{A}_L(H^{\otimes 2}) ||_{2}  + \frac{2^{n+2}}{2^{2n} - 1} \left( || \mathcal{A}_R(\rho^{\otimes 2}) ||_{2} (||H||_{2}^2-\frac{1}{d}\Tr[H]^2) +  || \mathcal{A}_L(H^{\otimes 2}) ||_{2} (||\rho||_{2}^2 - \frac{1}{d}) \right) \, . 
\end{align}

\medskip

\paragraph*{Extension to generalized cost.}
Similar to Eqs.~\eqref{eq:ar-rhojk-hjk} and \eqref{eq:al-rhojk-hjk}, the bound in Eq.~\eqref{eq:MainBoundLR} can be extended for the cost functions of the form in Eq.~\eqref{eq:gen-cost}. In particular, we find that 
\begin{align}
  | \text{Var} \,&\partial_k C - \text{Var}_{R, L} \partial_k C |   \nonumber \\ 
    &\leq 4\sum_{m,n}  || \mathcal{A}_{R}(\rho_m\otimes \rho_n) ||_{2} || \mathcal{A}_{L}(H_m\otimes H_n) ||_{2} \nonumber \\
    & \qquad + \frac{2^{n+2}}{2^{2n} - 1} \sum_{m,n}\left( || \mathcal{A}_{R}(\rho_m\otimes \rho_n) ||_{2} \bigg(\Tr[H_m H_n]-\frac{1}{d}\Tr[H_m]\Tr[H_n]\bigg) +  || \mathcal{A}_{L}(H_m\otimes H_n) ||_{2} \bigg(\Tr[\rho_m \rho_n] - \frac{1}{d}\bigg) \right) ~.
\end{align}

\subsection{Reformulating bounds using the diamond norm.} 
Here we derive bounds Eqs.~\eqref{eq:boundRd}-\eqref{eq:MainBoundLRd}, in which the expressiblity is quantified in terms of the diamond norm. This is a natural alternative way of formulating the bounds, since the diamond norm is an operationally meaningful measure of the distinguishability of two quantum operations that is often used to define $\varepsilon$-approximate $t$-designs.

To derive Eq.~\eqref{eq:boundRd} we start with Eq.~\eqref{eq:DiffVarR} and invoke the H\"{o}lder's inequality as follows: 
\begin{equation}
\begin{aligned}
        | \text{Var} \, \partial_k C - \text{Var}_R \partial_k C | &\leq \int_{\mathbb{U}_L} d U_L |  \Tr[\mathcal{A}_R(\rho^{\otimes 2}) X_{Lk}^{\otimes 2}] | \\
        &\leq  \int_{\mathbb{U}_L} d U_L || X_{Lk}^{\otimes 2}||_{\infty} || \mathcal{A}_R(\rho^{\otimes 2}) ||_1\, .
\end{aligned}
\end{equation}
The term $|| X_{Lk}^{\otimes 2}||_\infty$ can now be bounded as follows. Given that $X_{Lk}^\dagger = - X_{Lk}$, it follows from the unitary invariance and sub-multiplicativity of the infinity norm that
\begin{equation}
    || X_{Lk}^{\otimes 2}||_\infty = ( || X_{Lk}||_\infty )^2\, \leq ( 2 || V_k ||_\infty || U_L^\dagger H U_L ||_\infty )^2 = ( 2 || V_k ||_\infty || H ||_\infty )^2 \leq  4 || H ||_\infty^2
\end{equation}
We additionally note that $|| \mathcal{E}(X) ||_1 \leq  
\Vert X \Vert_1 || \mathcal{E} ||_{\diamond}$ for any channel $\mathcal{E}$ and operator $X$, therefore 
\begin{equation}
    || \mathcal{A}_R(\rho^{\otimes 2}) ||_1  \leq \Vert \rho \Vert_1 || \mathcal{A}_{U_R} ||_{\diamond} =  || \mathcal{A}_{U_R} ||_{\diamond} := \varepsilon^{\diamond}_R \, . 
\end{equation}
Thus we are now left with 
\begin{equation}
    | \text{Var} \, \partial_k C - \text{Var}_R \partial_k C | \leq  4 || H ||_\infty^2 \, \varepsilon^{\diamond}_R \, .
\end{equation}
The derivation of Eq.~\eqref{eq:boundLd} is entirely analogous. 

\medskip

To derive Eq.~\eqref{eq:MainBoundLRd} we start with Eq.~\eqref{eq:VarDiffLR} and again use H\"{o}lder's inequality in terms of the infinity and one norm to find 
\begin{equation}
    | \text{Var} \, \partial_k C - \text{Var}_{R, L} \partial_k C | \leq  \frac{1}{d^2 - 1 }((||\rho||_2^2-1/d) || Z_{Lk}||_1 + (||H||_2^2-\Tr[H]^2/d) || Z_{Rk}||_1) + || \mathcal{A}_R(\rho^{\otimes 2}) ||_{1} ||Z_{Lk}||_{\infty} \, ,
    \end{equation}
where we have used $|| W ||_{\infty} = 1$. 
Finally, by expanding $|| Z_{xk} ||_{\infty}$, using the triangle inequality and  and the fact that $V_k^2 = \I$ we find that
\begin{equation}
\begin{aligned}
        || Z_{xk} ||_{\infty} &\leq  || V_k^{\otimes 2} \mathcal{A}_x(\omega_x) ||_{\infty} + || \mathcal{A}_x(\omega_x) V_k^{\otimes 2} ||_{\infty} + 2 || (V_k \otimes \I) \mathcal{A}_x(\omega_x) (\I \otimes V_k)||_{\infty}  \\ 
        &\leq  2 || V_k^{\otimes 2}||_{\infty} || \mathcal{A}_x(\omega_x) ||_\infty + 2 || (V_k \otimes \I)  ||_\infty || \mathcal{A}_x(\omega_x) ||_{\infty} || (\I \otimes V_k)||_{\infty}  \\ 
        &\leq  4  || \mathcal{A}_x(\omega_x) ||_\infty  
\end{aligned} 
\end{equation}
for $x = L$ and $x = R$. We additionally note that $|| \mathcal{E}(X) ||_1 \leq \Vert X\Vert_1 || \mathcal{E} ||_{\diamond}$ for any channel $\mathcal{E}$ and operator $X$, therefore,  
\begin{align}
    || \mathcal{A}_R(\rho^{\otimes 2}) ||_{\infty}\leq || \mathcal{A}_R(\rho^{\otimes 2}) ||_1  \leq  \varepsilon^{\diamond}_R \\
   || \mathcal{A}_L(H^{\otimes 2}) ||_{\infty} \leq || \mathcal{A}_L(H^{\otimes 2}) ||_1  \leq \Vert H\Vert_1 \varepsilon^{\diamond}_L
\end{align}
Thus we are left with 
\begin{equation}
    | \text{Var} \, \partial_k C - \text{Var}_{R, L} \partial_k C | \leq  \frac{4}{d^2 - 1 }((||\rho||_2^2-1/d) || \mathcal{A}_L (H) ||_\infty  + (||H||_2^2-\Tr[H]^2/d) || \mathcal{A}_R(\rho^{\otimes 2}) ||_\infty ) + 4 || \mathcal{A}_R (\rho) ||_{1} || \mathcal{A}_L(H^{\otimes 2}) ||_\infty  \, 
\end{equation}
or alternatively 
\begin{equation}
    | \text{Var} \, \partial_k C - \text{Var}_{R, L} \partial_k C | \leq  \frac{4}{d^2 - 1 }((||\rho||_2^2-1/d) \Vert H\Vert_1 \eLd  + (||H||_2^2-\Tr[H]^2/d) \eRd ) + 4 \Vert H \Vert_1 \,  \varepsilon_{\scriptscriptstyle R}^{\scriptscriptstyle \diamond}  \varepsilon_{ \scriptscriptstyle L}^{\scriptscriptstyle \diamond}  \, . 
\end{equation}

\section{Variance in partial derivative for exact 2-designs.}\label{sec:google-high-order}
In this Appendix, we provide the explicit expressions and the derivation of the variance in the partial derivative for a random layered parametrized ansatz of the form Eqs.~\eqref{eq:LayeredAnsatz} and \eqref{eq:structure}--\eqref{eq:ulur}, and the generic cost defined in Eq.~\eqref{eq:GenCost}.
These quantities have been investigated in \cite{mcclean2018barren}; however, only the highest order terms in $n$ were given. Here we provide higher order terms for completeness. 

\medskip

\paragraph*{Explicit expressions}
Let us denote the variance of the cost when just $\mathbb{U}_R$, just $\mathbb{U}_L$, and both $\mathbb{U}_R$ and $\mathbb{U}_L$ form 2-designs as $\text{Var}_{R}\partial_k C $, $\text{Var}_{L}\partial_k C $, and $\text{Var}_{R, L}\partial_k C$, respectively. These variances are given by
\begin{align}
         & \text{Var}_x \partial_k C = \frac{g_x(\rho, H, U)}{2^{2n}-1} ~,  
\end{align}
where 
\begin{align}
    &g_R (\rho, H, U) = -\left(\Tr(\rho^2)-\frac{1}{2^n}\right)\int dU_L\Tr([V_k, U_L^{\dagger}HU_L]^2) \\ 
    &g_L (\rho, H, U) = - \left(\Tr(H^2) - \frac{\Tr[H]^2}{2^n}\right)\int dU_R \Tr([V_k, U_R^{\dagger} H U_R]^2) \\
    &g_{R, L} (\rho, H, U) = -2 \Big(\Tr(\rho^2)-\frac{1}{2^n}\Big) \Big(\frac{1}{2^{2n}-1}[\Tr(V_k)^2 \Tr(H^2) + \Tr(V_k^2)\Tr(H)^2]\nonumber \\
  &\qquad \ \ \ \ \ \ \ \ \ \ \ \ \ \ \ 
  - \frac{1}{2^n(2^{2n}-1)}[\Tr(V_k^2)\Tr(H^2)+\Tr(V_k)^2\Tr(H)^2]- \frac{1}{2^n}\Tr(V_k^2) \Tr(H^2)\Big)
\end{align}

\paragraph*{Derivation}
From Eq.~\eqref{eq:gradCost}, we have
\begin{equation}
    \partial_k C := \frac{\partial C}{\partial \theta_k } = i  \Tr[ U_R \rho U_R^\dagger [V_k, U_L^\dagger H U_L] ] \, . 
\end{equation}

Since the cost gradient is unbiased, as in Eq.~\eqref{eq:GradUnbiased}, the variance in the partial derivative is given by
\begin{align}\label{eq:var-general-form}
    \text{Var}\partial_k C &= - \int dU_L \int dU_R  \Tr(U_R  \rho U_R^{\dagger}[V_k, U_L^{\dagger}HU_L] )^2.
\end{align}    

Then $\text{Var}_R\partial_k C$, $\text{Var}_L\partial_k C$, and $\text{Var}_{R,L}\partial_k C$ can be calculated by the integration in Eq.~\eqref{eq:var-general-form}  over $U_R$, $U_L$, and both $U_R$ and $U_L$, respectively. 

Integrating over only $U_R$ gives 
\begin{align}
\text{Var}_R\partial_k C    & = - \frac{1}{d^2 -1}\int dU_L (\Tr(\rho)^2\Tr([V_k, U_L^{\dagger}HU_L])^2 + \Tr(\rho^2)\Tr([V_k, U_L^{\dagger}HU_L]^2))  \nonumber \\
    & \qquad +\frac{1}{d(d^2-1)}\int dU_L  (\Tr(\rho^2)\Tr([V_k, U_L^{\dagger}HU_L])^2 + \Tr(\rho)^2 \Tr([V_k, U_L^{\dagger}HU_L]^2) ) \label{eq:varRderivation1} \\
    & = - \frac{1}{d^2-1} \int dU_L \Tr(
    \rho^2) \Tr([V_k, U_L^{\dagger}HU_L]^2) + \frac{1}{d(d^2-1)} \int dU_L\Tr([V_k, U_L^{\dagger}HU_L]^2)\label{eq:varRderivation2}\\
    & = -\left(\Tr(\rho^2)-\frac{1}{d}\right)\left(\frac{1}{d^2-1}\right)\int dU_L\Tr([V_k, U_L^{\dagger}HU_L]^2)~,\label{eq:varRderivation3}
\end{align}
where the first equality follows from Eq.~\eqref{eq:identity3} and the second equality follows from the fact that the trace of a commutator is always zero.

Form the cyclicity of the trace operation and the arguments similar to Eqs.~\eqref{eq:varRderivation1} and \eqref{eq:varRderivation2}, we get 
\begin{align}
    \text{Var}_L\partial_k C = - \left(\Tr(H^2) - \frac{\Tr[H]^2}{d}\right)\left(\frac{1}{d^2-1}\right) \int dU_R \Tr([V_k, U_R^{\dagger} H U_R]^2)~.
\end{align}

In order to calculate $\text{Var}_{R,L}\partial_k C$, we note that $\Tr([V_k, U_L^{\dagger}HU_L]^2)$ in Eq.~\eqref{eq:varRderivation3} can be written as 
\begin{align}
    \Tr([V_k, U_L^{\dagger}HU_L]^2) = 2[\Tr(U_LV_k U_L^{\dagger}& H U_L V_k U_L^{\dagger} H ) - \Tr(U_L V_k^2 U_L^{\dagger} H^2)]~.\label{eq:final-int-lr}
\end{align}

The integral of the first term over $U_L$ in Eq.~\eqref{eq:final-int-lr} can be calculated using Eq.~\eqref{eq:identity2} as follows: 
\begin{align}
    \int dU_L \Tr(U_LV_k U_L^{\dagger}& H U_L V_k U_L^{\dagger} H ) \nonumber  \\
   &= \frac{1}{d^2-1}[\Tr(V_k)^2 \Tr(H^2) + \Tr(V_k^2)\Tr(H)^2] - \frac{1}{d(d^2 -1)}[\Tr(V_k^2)\Tr(H^2)+\Tr(V_k)^2\Tr(H)^2]. 
\end{align}

The integral of the second term in Eq.~\eqref{eq:final-int-lr} can be calculated using Eq.~\eqref{eq:identity1} as follows:
\begin{align}
\int dU_L \Tr[U_LV_k^2U_L^{\dagger}H^2] = \frac{\Tr(V_k^2)\Tr(H^2)}{d}~.
\end{align}

Finally, after combining everything we get 
\begin{align}
  \text{Var}_{R,L}\partial_kC &=  -\Big(\Tr(\rho^2)-\frac{1}{d}\Big)\Big(\frac{2}{d^2-1}\Big) \Big(\frac{1}{d^2-1}[\Tr(V_k)^2 \Tr(H^2) + \Tr(V_k^2)\Tr(H)^2]\nonumber \\
  &\qquad
  - \frac{1}{d(d^2 -1)}[\Tr(V_k^2)\Tr(H^2)+\Tr(V_k)^2\Tr(H)^2]- \frac{1}{d}\Tr(V_k^2) \Tr(H^2)\Big)
\end{align}

\section{Concentration of measure}\label{ap:Conc}

In Ref.~\cite{marrero2020entanglement} it was shown that for ans\"{a}tze where the reduced state on the measured qubits obeys a volume law, typical local cost function values concentrate exponentially fast in $n$ to its mean. This result was complemented by a proof that for ans\"{a}tze that form 2-designs, i.e. maximally expressive ans\"{a}tze, local costs concentrate exponentially fast to a fixed value. Here we show that this proof may be generalised to non-perfectly expressive ans\"{a}tze.

Specifically we show that for a $k$-local cost $C_k$ we have that 
\begin{align}\label{eq:concentrationresult}
 \langle  | C_{k} - \Tr( ( H_k \otimes \I ) \I/d ) |\rangle \leq || H_k ||_\infty \left( \sqrt{  \int dU_{\rm Haar}  \Tr(  ( U\otimes U ) (\sigma \otimes \sigma)  ( U^\dagger \otimes U^\dagger )  (  W \otimes \I )) )}\right) + || H_k ||_\infty \sqrt{\chi_\epsilon }~,
\end{align}
where
\begin{equation}\label{eq:concentrationofmeasure}
    \left( \sqrt{  \int dU_{\rm Haar}  \Tr(  ( U\otimes U ) (\sigma \otimes \sigma)  ( U^\dagger \otimes U^\dagger )  (  W \otimes \I )) )}\right)   \in \mathcal{O}\left( \sqrt{\frac{2^k}{2^n}}\right)~.
\end{equation}
 Here $\chi_\epsilon$ is an expressibility dependent correction defined as 
\begin{equation}
   \chi_\epsilon  := \Tr(  \mathcal{A}_{\mathbb{U}}(|0\rangle \langle 0 | ) (  W \otimes \I )) )  \, , 
\end{equation} where, as previously, $W$ is the subsystem permuation operator.

\begin{proof}
The start of the proof of is identical to Ref.~\cite{marrero2020entanglement}. 
\begin{equation}
\begin{aligned}
    \langle  | C_{k} - \Tr( ( H_k \otimes \I ) \I/d ) | \rangle &= \int dU \vert( ( H_k \otimes \I ) \left( U |0\rangle \langle 0 | U^\dagger -  \I/d \right) ) | \\
   &\leq || H_k ||_\infty \int dU ||\Tr_{\overline{k}}(  \left( U |0\rangle \langle 0 | U^\dagger -  \I/d \right) ) ||_1 \\
  &\leq || H_k ||_\infty \sqrt{ 2^{k} \int dU || \Tr_{\overline{k}}(  \left( U |0\rangle \langle 0 | U^\dagger -  \I/d \right) ) ||_2^2 } \\ 
  &= || H_k ||_\infty \sqrt{  \int dU  \Tr( \left( U |0\rangle \langle 0 | U^\dagger -  \I/d \right) \otimes  \left( U |0\rangle \langle 0 | U^\dagger -  \I/d \right) (  W \otimes \I )) )  } \\
   &= || H_k ||_\infty \sqrt{  \int dU  \Tr(  ( U\otimes U ) (\sigma \otimes \sigma)  ( U^\dagger \otimes U^\dagger )  (  W \otimes \I )) )  }
\end{aligned}
\end{equation}
where $\sigma = |0\rangle \langle 0 | -  \I/d$. The first inequality follows from H\"{o}lder's inequality. For the second inequality, we used the relation between the trace norm and the Hilbert-Schmidt norm and invoked Jensen's inequality. We used $\overline{k}$ to denote qubits that are not measured for defining the cost function $C_k$.

We now substitute in the definition of $\mathcal{A}_{\mathbb{U}}(|0\rangle \langle 0 | )$ to get that 
\begin{equation}
\int dU   \Tr(  ( U\otimes U ) (\sigma \otimes \sigma)  ( U^\dagger \otimes U^\dagger )  (  W \otimes \I )) ) = \int dU_{\rm Haar}   \Tr(  ( U\otimes U ) (\sigma \otimes \sigma)  ( U^\dagger \otimes U^\dagger )  (  W \otimes \I )) ) +   \Tr(  \mathcal{A}_{\mathbb{U}}(|0\rangle \langle 0 | ) (  W \otimes \I )) )  \, .
\end{equation}
Introducing the short hand $  \Tr(  \mathcal{A}_{\mathbb{U}}(|0\rangle \langle 0 | ) (  W \otimes \I )) ) = \chi_\epsilon$ to denote the expressibility dependent correction we can then write 
\begin{equation}
\begin{aligned}
     \langle  | C_{k} - \Tr( ( H_k \otimes \I ) \I/d ) | \rangle  &\leq || H_k ||_\infty \sqrt{  \int dU_{\rm Haar}  \Tr(  ( U\otimes U ) (\sigma \otimes \sigma)  ( U^\dagger \otimes U^\dagger )  (  W \otimes \I )) ) + \chi_\epsilon } \\
    &\leq || H_k ||_\infty \left( \sqrt{  \int dU_{\rm Haar}  \Tr(  ( U\otimes U ) (\sigma \otimes \sigma)  ( U^\dagger \otimes U^\dagger )  (  W \otimes \I )) )} + \sqrt{\chi_\epsilon } \right)
\end{aligned}
\end{equation}
where we have used $\sqrt{a + b } \leq \sqrt{a} + \sqrt{b}$. Moreover, Eq.~\ref{eq:concentrationofmeasure} follows from Theorem~2 in \cite{popescu2005foundations}, which completes the proof.  
\end{proof}

\section{Numerically studying the correlations between expressibility and cost partial derivatives}\label{ap:ExpressVersusGrads}

In this Appendix we present numerical results on the correlations between the cost gradient and expressibility. Specifically, we consider the layered parametrized ansatz detailed in Section~\ref{sec:Numerics} of the main text and plot the variance in the PQC gradients as a function of its expressibility. 

We can calculate the expressibility measures $\varepsilon_R^\rho$ and $\varepsilon_L^H$ via their reformulation in terms of the state and Hamiltonian dependent frame potentials $\mathcal{F}_R^{(\rho)} := \mathcal{F}_{\mathbb{U}_R}^{(\rho)}$ and $\mathcal{F}_L^{(H)} := \mathcal{F}_{\mathbb{U}_L}^{(\rho)}$ given in Eq.~\eqref{eq:FPs}. However, since it follows from Eq.~\eqref{eq:FPsHaar} that the state (Hamiltonian) dependent frame potential for the Haar distribution $\mathcal{F}^{(\rho)}_{\rm Haar}$ ($\mathcal{F}^{(H)}_{\rm Haar}$) is exponentially small, and $\varepsilon_{R}^\rho$ ($\varepsilon_{L}^H$) measures the difference between $\mathcal{F}_R^{(\rho)}$ and $\mathcal{F}^{(\rho)}_{
\rm Haar}$ ($\mathcal{F}_L^{(H)}$ and $\mathcal{F}^{(H)}_{
\rm Haar}$), it follows that $\varepsilon_{R}^\rho$ ($\varepsilon_{L}^H$) may also be exponentially small. We therefore find the ratio of the true frame potential to the Haar frame potential more insightful to plot. That is, we consider the ratios
\begin{align}
    &\frac{\mathcal{F}_R^{(\rho)}}{\mathcal{F}^{(\rho)}_{\rm Haar}} = \frac{(\varepsilon_U^\rho)^2}{\mathcal{F}^{(\rho)}_{\rm Haar}} + 1     \\
    &\frac{\mathcal{F}_L^{(H)}}{\mathcal{F}^{(H)}_{\rm Haar}} = \frac{(\varepsilon_U^H)^2}{\mathcal{F}^{(H)}_{\rm Haar}} + 1 \, .
\end{align}
The larger these ratios, the more inexpressive the ansatz, with the ratios tending to 1 for maximally expressive ans\"{a}tze (exact 2-designs).

\begin{figure*}[t!]
    \centering
    \includegraphics[width=\textwidth]{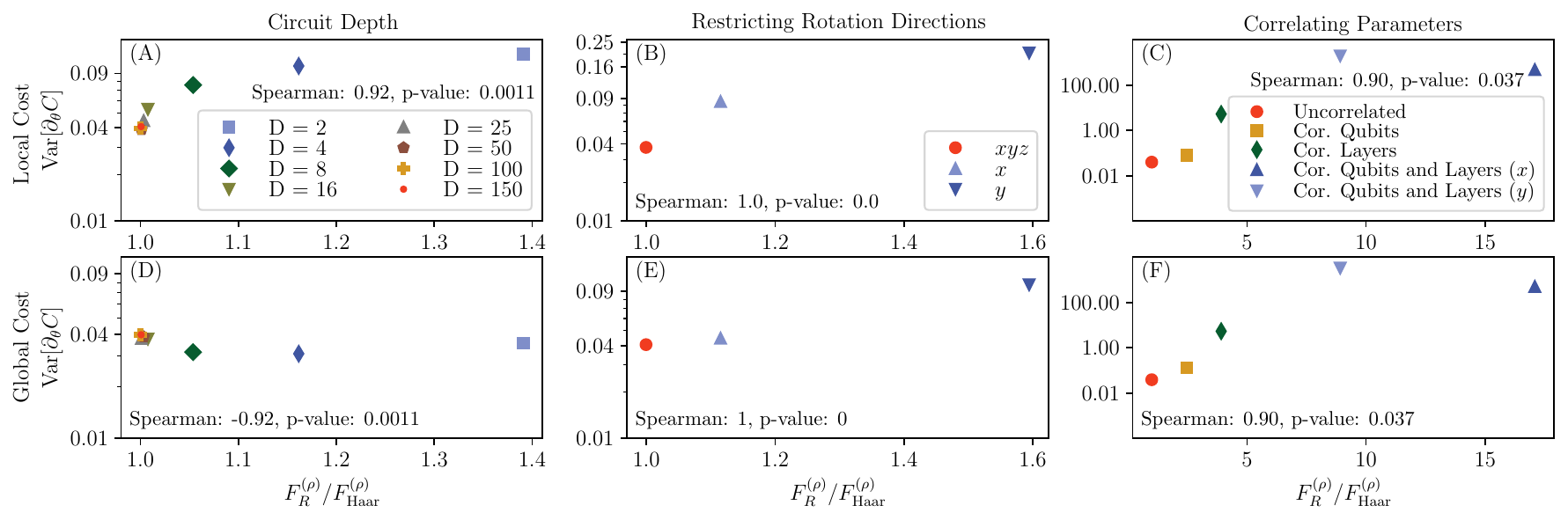}
    \caption{\textbf{Correlations between cost partial derivatives and the \textit{state}-dependent frame potential.} The variance in the partial derivative of a global cost with $H = \prod_{i=1}^n \sigma^z_i$ (top) and 2-local cost with $H = \sigma^z_1 \sigma^z_2$ (bottom) as a function of the expressibility measure $\frac{\mathcal{F}_R^{(\rho)}}{\mathcal{F}^{(\rho)}_{\rm Haar}}$ (in both cases $\rho = |\psi_0\rangle \langle \psi_0 |^{\otimes n} $ where $\ket{\psi_0} = \exp(-i(\pi/8)\sigma_Y) \ket{0}$). In the left panel we vary the circuit depth $D$ of a hardware efficient ansatz. In the right (middle) panel we consider the effect of correlating parameters (restricting the directions of rotation) of a hardware efficient ansatz with $D =100$ with the choices of correlations (rotations) indicated in the figure legend. In all cases $n=4$, the derivative is taken with respect to $\theta_D^1$ and the variance and frame potentials are estimated using an ensemble of $5000$ unitaries.}\label{fig:ExpressVersusGradsRho}
\end{figure*}

\begin{figure*}[t!]
    \centering
    \includegraphics[width=\textwidth]{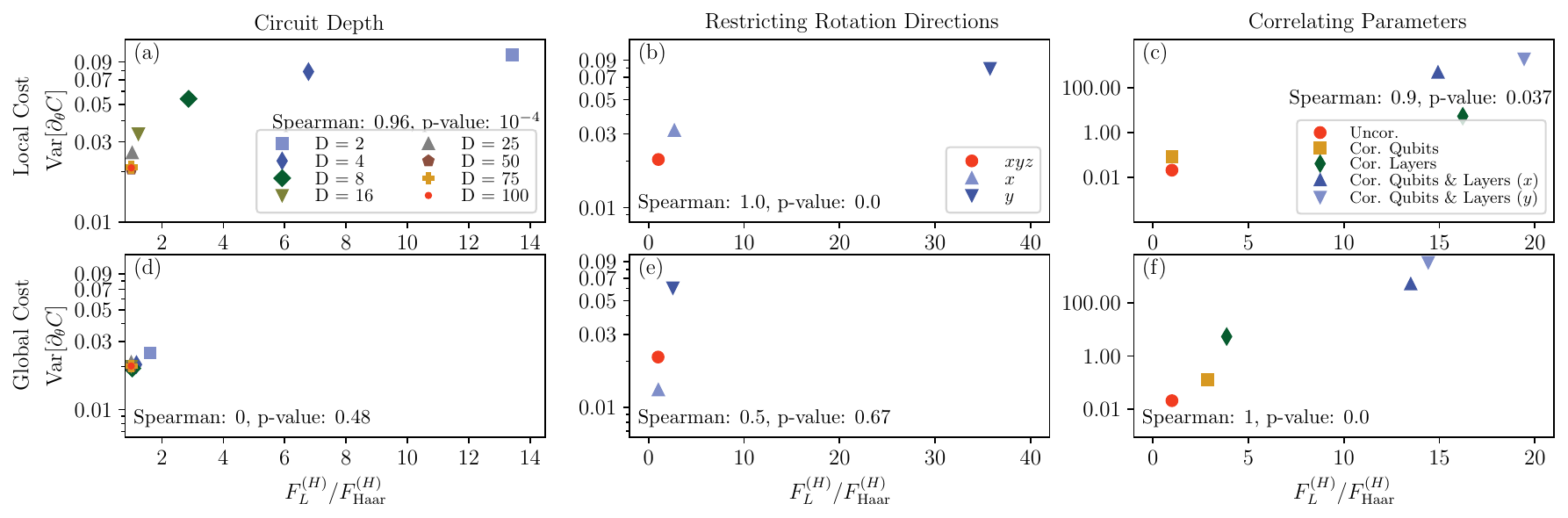}
    \caption{\textbf{Correlations between cost partial derivative and the \textit{Hamiltonian}-dependent frame potential.} This setting here is entirely equivalent to that described in Fig.~\ref{fig:ExpressVersusGradsRho}; however, here we plot the variance in the partial derivative as a function of the ratio of \textit{Hamiltonian}-dependent frame potentials $\frac{\mathcal{F}_L^{(H)}}{\mathcal{F}^{(H)}_{\rm Haar}}$ and the derivative is taken with respect to $\theta_1^1$.}\label{fig:ExpressVersusGradsHam}
\end{figure*}

In Fig.~\ref{fig:ExpressVersusGradsRho} and Fig.~\ref{fig:ExpressVersusGradsHam} we plot the variance in the partial derivative as a function of $\frac{\mathcal{F}_R^{(\rho)}}{\mathcal{F}^{(\rho)}_{\rm Haar}}$ and $\frac{\mathcal{F}_L^{(H)}}{\mathcal{F}^{(H)}_{\rm Haar}}$ respectively.
Inline with Section~\ref{sec:Numerics} of the main text, we focus on three different ways of tuning the expressibility of an ansatz; namely decreasing the depth of the circuits, correlating circuit parameters, and restricting either the direction of rotations. 

To numerically quantify the degree of correlations between the variance in the partial derivative of the cost and the expressibility we include in Fig.~\ref{fig:ExpressVersusGradsRho} and Fig.~\ref{fig:ExpressVersusGradsHam} the Spearman correlation coefficient~\cite{choi1977tests} and its corresponding p-value. Overall we find a clear correlation between partial derivatives of the cost and expressibility, with the variance in the derivatives increasing with increasing $\mathcal{F}_R^{(\rho)}$ and $\mathcal{F}_L^{(H)}$. Specifically, combining all the different ways of tuning the expressibility, the Spearman coefficient for the correlation between the variance in the partial derivative of the cost and the $\epsilon^\rho$ was found to be 0.78 with a p-value of $1.19 \times 10^{-7}$. Similarly, for the correlations with $\epsilon^H$ was 0.80 with a p-value of $1.18 \times 10^{-7}$.

It is noteworthy that the Hamiltonian dependent frame potential captures the effect of locality on cost gradients as the circuit depth is tuned. As observed in Section~\ref{sec:Numerics}, increasing the depth of the circuit reduces cost partial derivatives for a local cost but not a global cost. The state dependent frame potential cannot capture this effect since it is independent of the choice in measurement operator $H$ and therefore necessarily independent of the locality of $H$. Conversely, the while the Hamiltonian frame potential for a local cost decreases with increasing depth, inline with the decreasing variance in partial derivatives, the Hamiltonian dependent frame potential for the global cost is effectively constant (even as the depth of the circuit substantially increases) reflecting the effectively constant variance in partial derivatives.

Nonetheless, the correlation between the variance in the cost partial derivative and the expressibility is not perfect, as is clear for example, from Fig.~\ref{fig:ExpressVersusGradsHam}(F). This is entirely compatible with our analytical bounds, which are upper bounds and therefore do not enforce perfect correlation between the variance in the partial derivative and the expressibility. Thus while Fig.~\ref{fig:ExpressVersusGradsRho} and Fig.~\ref{fig:ExpressVersusGradsHam} demonstrate a clear correlation between the variance in the partial derivative of the cost and expressibility, further work is required to understand the intricacies of this correlation.

\end{document}